\documentclass[fleqn,usenatbib]{mnras}

\usepackage{newtxtext,newtxmath}
\usepackage[normalem]{ulem}

\usepackage[T1]{fontenc}

\DeclareRobustCommand{\VAN}[3]{#2}
\let\VANthebibliography\thebibliography
\def\thebibliography{\DeclareRobustCommand{\VAN}[3]{##3}\VANthebibliography}

\def\alstar{{\textsc a}l{\textsc s}tar}
\def\starlight{\textsc{starlight}}
\newcommand{\hii}{H\thinspace\textsc{ii}}
\newcommand{\Ha}{\ifmmode \mathrm{H}\alpha \else H$\alpha$\fi}
\newcommand{\Hb}{\ifmmode \mathrm{H}\beta \else H$\beta$\fi}
\newcommand{\nii}{\ifmmode [\mathrm{N}\,\textsc{ii}] \else [N~{\scshape ii}]\fi}
\newcommand{\oiii}{\ifmmode [\mathrm{O}\,\textsc{iii}] \else [O\,{\scshape iii}]\fi}
\newcommand{\sii}{\ifmmode [\mathrm{S}\,\textsc{ii}] \else [S~{\scshape ii}]\fi}
\newcommand{\oi}{\ifmmode [\mathrm{O}\,\textsc{i}] \else [O\,{\scshape i}]\fi}
\newcommand{\oii}{\ifmmode [\mathrm{O}\,\textsc{ii}] \else [O\,{\scshape ii}]\fi}
\newcommand{\nOne}{\ifmmode [\mathrm{N}\,\textsc{i}] \else [N~{\scshape i}]\fi}
\newcommand{\siii}{\ifmmode [\mathrm{S}\,\textsc{iii}] \else [S~{\scshape iii}]\fi}

\newcommand{\HaNii}{\ifmmode \mathrm{H}\alpha\mathrm{N}\textsc{ii} \else H$\alpha$N{\scshape ii}\fi}
\newcommand{\WHaNii}{\ifmmode \mathrm{W}_{\mathrm{H}\alpha\mathrm{N}\textsc{ii}} \else W$_{\mathrm{H}\alpha\mathrm{N}\textsc{ii}}$\fi}

\newcommand{\WHaN}{\ifmmode W_{{\mathrm H}\alpha{\mathrm N}} \else $W_{{\rm H}\alpha{\rm N}}$}
\newcommand{\WHa}{\ifmmode W_{{\rm H}\alpha} \else $W_{{\rm H}\alpha}$}

\usepackage[usanames]{xcolor}
\definecolor{Jpurple}{RGB}{ 155, 0, 255 }

\usepackage{graphicx}	
\usepackage{amsmath}	
\usepackage{multirow}

\title[Stellar population and emission line properties in S-PLUS galaxies]{Estimating stellar population and emission line properties in S-PLUS galaxies}

\author[J. Thainá-Batista et al.]{
J. Thainá-Batista$^{1}$\thanks{E-mail: jullia.thainna@gmail.com}, 
R. Cid Fernandes$^{1}$, 
F. R. Herpich$^{2,3}$, 
C. Mendes de Oliveira$^{2}$, 
A. Werle$^{4}$, 
L. Espinosa$^{2}$, \newauthor
A. Lopes$^5$, 
A. V. Smith Castelli$^{5,6}$, 
L. Sodré$^{2}$,
E. Telles$^7$,
A. Kanaan$^1$,
T. Ribeiro$^8$, 
W. Schoenell$^9$
\\
$^{1}$Departamento de Física - CFM - Universidade Federal de Santa Catarina, PO BOx 476, 88040-900, Florianópolis, SC, Brazil \\
$^{2}$Departamento de Astronomia, Instituto de Astronomia, Geofísica e Ciências Atmosféricas da USP, Cidade Universitária, 05508-900, São Paulo, SP, Brazil\\
$^{3}$Cambridge Astronomy Survey Unit, Institute of Astronomy, Madingley Road, Cambridge CB3 0HA, UK\\
$^{4}$INAF—Osservatorio Astronomico di Padova, vicolo dell’Osservatorio 5, I-35122 Padova, Italy\\
$^{5}$Instituto de Astrofsica de La Plata, CONICET–UNLP, Paseo del Bosque s/n, B1900FWA, Argentina\\
$^{6}$Facultad de Ciencias Astronómicas y Geofísicas, Universidad Nacional de La Plata, Paseo del Bosque s/n, B1900FWA, Argentina\\
$^{7}$Observatório Nacional, Rua General José Cristino 77, CEP: 20921-400, São Cristóvão, Rio de Janeiro, Brasil\\
$^{8}$NOAO, 950 North Cherry Ave. Tucson, AZ 85719, United States\\
$^{9}$GMTO Corporation, N. Halstead Street 465, Suite 250, Pasadena, CA 91107, United
States
}

\date{Accepted XXX. Received YYY; in original form ZZZ}

\pubyear{2023}

\begin{document}
\label{firstpage}
\pagerange{\pageref{firstpage}--\pageref{lastpage}}
\maketitle
 
\begin{abstract} 
We present tests of a new  method to simultaneously estimate stellar population and emission line (EL) properties of galaxies out of S-PLUS photometry. The technique uses the \alstar\ code, updated with an empirical prior which greatly improves its ability to estimate ELs using only the survey's 12 bands. The tests compare the output of (noise-perturbed) synthetic photometry of SDSS galaxies to properties derived from previous full spectral fitting and detailed EL analysis.
For realistic signal-to-noise ratios, stellar population properties are recovered to better than 0.2 dex in masses, mean ages, metallicities and $\pm 0.2$ mag for the extinction. More importantly, ELs are recovered  remarkably well for a photometric survey. 
We obtain input $-$ output dispersions of 0.05--0.2 dex for the equivalent widths of \oii, \oiii, \Hb, \Ha, \nii, and \sii, and even better for lines stronger than $\sim 5$ \AA. 
These excellent results are achieved by combining two empirical facts into a prior which restricts the EL space available for the fits:
(1) Because, for the redshifts explored here, \Ha\ and \nii\ fall in a single narrow band (J0660), their combined equivalent width is always well recovered, even when \nii/\Ha\ is not. (2) We know from SDSS that $W_{\Ha+\nii}$ correlates with \nii/\Ha, which can be used to tell if a galaxy belongs to the left or right wings in the classical BPT diagnostic diagram. Example applications to integrated light and spatially resolved data are also presented, including a comparison with independent results obtained with MUSE-based integral field spectroscopy.

\end{abstract}




\section{Introduction}

The Southern Photometric Local Universe Survey (S-PLUS) is an ongoing project based on an 80 cm robotic telescope located at Cerro Tololo (Chile), which gathers images on five broad and seven narrow bands spanning the $\sim 3500$--9000 \AA\ range. Like its northern twin, the Javalambre Photometric Local Universe Survey (J-PLUS; \citealt{jplus}), the S-PLUS is a multi-purpose survey, with applications ranging from Solar system to extragalactic scales. 
A full description of S-PLUS is given in \cite{2019Splus}. Some of its first results are reported in \citet{Barbosa-2020}, \citet{Molino-2020}, \citet{Lima-Dias-2020}, \citet{Whitten-2021}, \citealt{Placco-2021}, \citet{Nakazono-2021}, \citet{LIMA2022100510} and \citet{Almeida-Fernandes-2022}.
 
This paper addresses the question of how to use S-PLUS data to characterise the basic stellar population and emission line (EL) properties of galaxies. Estimates of properties like stellar masses, mean stellar ages and emission line equivalent widths are to study a variety of science cases in galaxy evolution, like those involving galaxy morphology or environment, as well as in the spatially resolved analysis of nearby galaxies. The literature is plentiful on methods and tools to fit the spectral-energy-distribution (SED) of galaxies, both for spectroscopic and photometric data \citep[see][]{Conroy2013, Carnall2019, Leja2019}. Few, however, have been tested with the S-PLUS filter system, whose peculiar combination of narrow and broad bands calls for simultaneous analysis of stellar and nebular emission components. 
\cite{2015_Vilella_Rojo} and \cite{2019Logrono_Garcia} did explore this issue in the context of J-PLUS, but focusing on the estimation of the \Ha\ fluxes.

We present a series of experiments with the \alstar\ code, applied to both simulated and actual S-PLUS data. \alstar\ was  first presented in \citet[hereafter GD21]{JPAS_2021_delgado}
in a study of galaxies in the miniJPAS survey \citep{2021Bonoli}. Besides the very different number of bands (12 vs. \ 56), in GD21, all bands potentially contaminated by \oii3726,3729, \Hb, \oiii4959,5007, \Ha, and \nii6548,6584 were discarded from the analysis, whereas here we do not perform such masking. 
Instead, we simultaneously model both stellar and nebular emissions using a semi-empirical approach which ensures that the resulting ELs are realistic.

The paper is organised as follows. Section \ref{sec:AlStar} describes the \alstar\ code. Emphasis is given to how we account for ELs, as this is the most innovative aspect of this study. Section \ref{sec:SDSS_simulations} presents a series of simulations to evaluate the ability of the code to retrieve stellar and nebular properties out of S-PLUS photometry under different noise levels. The input in these simulations is based on SDSS spectra whose stellar populations and ELs have been previously analysed by \citet[hereafter W19]{Werle_2019} with the \starlight\ \citep{Cid2005} and {\sc dobby} codes \citep{2019Vale_Asari}.
Section \ref{sec:Applications} presents a few example applications to actual S-PLUS data, including the analysis a data cube of a nearby spiral galaxy. Finally, Section \ref{sec:Conclusions} summarises our main results.

\section{Spectral synthesis}
\label{sec:AlStar}

The analysis of galaxy SEDs employing  fits with stellar population models  dates back to Tinsley and others in the 1970s (see \citealt{Walcher2011} and \citealt{Conroy2013} for reviews). This spectral synthesis approach, as it is sometimes called, has progressed substantially in the last two decades with extensive work on evolutionary tracks and libraries of stellar spectra, resulting in improved models for stellar populations of different ages and metallicities. These key ingredients are used in codes which mix these populations (following either parametric or non-parametric prescriptions) to estimate properties such as stellar masses, mean stellar ages and metallicities, and dust content by comparing the models to spectroscopic (e.g., \citealt{2007Asari}, \citealt{2009_Riffel}) or photometric data \citep[e.g.][]{MAGPHYS, CIGALE, BEAGLE, BAGPIPES}.

The \alstar\ code described in this section performs a non-parametric decomposition of the input photometric (and spectroscopic, when available) fluxes in terms of a spectral base composed of stellar populations and, optionally, ELs. The code was introduced and compared with three other  codes in GD21 in an analysis of $\sim 8500$ galaxies in the AEGIS field (the miniJPAS survey; \citealt{2021Bonoli}). The data comprised photometry covering the $\sim 3500$--9300 \AA\ range obtained with 54 narrow band filters (FWHM $\sim 145$ \AA) spaced by $\sim 100$ \AA, plus two broader bands at the blue and red ends. 
The analysis focused entirely on the stellar population properties retrieved by the different codes. The potential effects of nebular emission on the photometry were circumvented by removing all bands covering the main optical ELs from the fits. The performance of \alstar\ in dealing with ELs was therefore {\em not} tested in GD21.

While suitable for spectroscopy (or when tens of narrow bands are available), the strategy of {\em (i)} masking ELs, {\em (ii)} fitting the stellar continuum on the unmasked data, and {\em (iii)} measuring ELs from the residual spectrum, is clearly not an optimal work-flow in the case of S-PLUS. First, ELs are present in all bands, so neglecting their effect in some filters (say, the broad bands) already limits the precision of the stellar continuum fits. Secondly, the amount of EL information retrievable from such a residual photo-spectrum would be very limited, with no guarantee that the inferred line fluxes are realistic.

This paper showcases our method to account simultaneously for stellar and nebular emission in \alstar. The method applies to any set of filters, but we focus on the 12 S-PLUS bands. We start by summarising the base of stellar population models employed and how dust attenuation is modelled (Section \ref{sec:StellarBaseAndDust}). We then present our semi-empirical approach to account for ELs (Section \ref{sec:NebularBase}) and how we tune it to improve the results (Section \ref{sec:WHaN_based_constraints}).

\subsection{Stellar population base and dust attenuation}
\label{sec:StellarBaseAndDust}

The main stellar population base used in this work is the same as in GD21 and very similar to the one in W19. Briefly, it contains spectra for 16 $\sim$ logarithmically spaced age bins spanning from $t = 1$ Myr to 14 Gyr, and seven metallicities from $Z = 0.005$ to $3.5 Z_\odot$, built out of an updated version of the \cite{Bruzual_charlot2003} models (see \citealt{2017Vidal_Garcia} and W19 for details on the evolutionary tracks and spectral libraries involved). A \cite{Chabrier_2003} initial mass function is adopted.

Dust attenuation is parameterised by the $V$-band optical depth ($\tau$) and is modelled with a \cite{2000Calzetti} law in all cases in this paper. \alstar\ allows for up to two different values of $\tau$: one applied to the full base ($\tau_{\rm ISM}$) and an extra one ($\tau_{\rm BC}$) to just some components. 
The motivation is to allow young stars to suffer an extra attenuation due to dust in their surrounding birth clouds (the so-called differential-extinction phenomenon, \citealt{cks94} and \citealt{Charlot_2000}). The two $\tau$'s can be either free or tied together. We chose to set $\tau_{\rm BC} = 1.27 \tau_{\rm ISM}$ for $\le 10$ Myr stars (as well as ELs) so that these components are 2.27 times as attenuated as older populations (cf. \citealt{2001Calzetti}). Other configurations are possible, including some where $\tau_{\rm BC}$ is not necessarily associated with birth clouds, but these will not be explored in this work.

After fitting the photometry, \alstar\ outputs how much each base component contributes to the flux at a chosen reference rest-frame wavelength of 5635 \AA, from which other properties like stellar masses and (light or mass-weighted) mean ages can be readily computed. To put it in mathematical terms, the model for the stellar spectrum reads:

\begin{equation}
\label{eq:M_star_lambda}
M^\star_\lambda = \sum_{j=1}^{n_\star} x_j 
b^\star_{\lambda,j} e ^{-\tau_j q_\lambda},
\end{equation}

\noindent where $b^\star_{\lambda,j}$ is the spectrum of population $j$ (age $t_j$ and metallicity $Z_j$) scaled to 1 at $\lambda = 5635$ \AA, $q_\lambda = \tau_\lambda/\tau_V$ is the reddening law, and $\tau_j = \tau_{\rm ISM}$ for $t_j > 10$ Myr or $\tau_j = \tau_{\rm ISM} + \tau_{\rm BC} = 2.27 \tau_{\rm ISM}$ for younger populations. 
The parameters in this model are the $x_j$ fluxes of the $n_\star$ populations and $\tau_{\rm ISM}$, while $b^\star_{\lambda,j}$ and $q_\lambda$ are its ingredients.

The fit is repeated $n_{\rm MC} = 100$ times perturbing the input fluxes with the corresponding errors. These Monte Carlo (MC) runs map both noise-induced uncertainties and the degeneracies inherent to spectral synthesis.

\subsection{The emission-line base}
\label{sec:NebularBase}

ELs are accounted for with a novel approach which incorporates them in the spectral base in a constrained manner. Five sets of lines are considered: the \oii$\lambda\lambda3726,3729$ doublet, \oiii$\lambda\lambda4959,5007$, \nii$\lambda\lambda6548,6584$, \sii$\lambda\lambda6717,6731$, and the Balmer series (from \Ha\ to H$\epsilon$).\footnote{We denote \oii$\lambda\lambda3726+3729$, \oiii$\lambda5007$, \nii$\lambda6584$, \sii$\lambda\lambda6716+6731$ by \oii, \oiii, \nii, and \sii, respectively. 
The combination of \Ha\ with both the \nii$\lambda\lambda6548,6584$ lines is denoted by 
\HaNii.
}
Each of these sets has relative line intensity ratios pre-defined by  nebular physics. The \oiii5007/\oiii4959 and \nii6584/\nii6548 flux ratios are fixed at 3, while the relative strengths of the \sii\ lines are fixed at its $\sii6717/\sii6731 = 1.4$ low-density limit, and similarly for 
\oii3726/\oii3729.
The relative strengths of Balmer lines are fixed at the values obtained for an \hii\ region with electron temperature $= 10^4$ K and density $= 10^2$ cm$^{-3}$, as given in the emissivity tables of \cite{2003Dopita_Sutherland}.

\begin{figure}
  \includegraphics[width=\columnwidth]{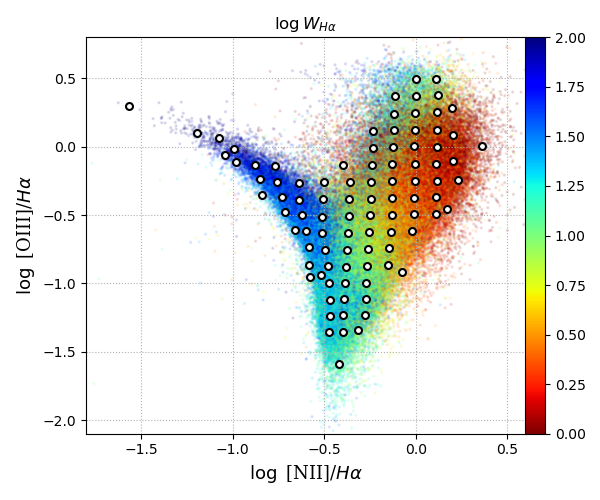}
  \caption{The BPT$\alpha$ diagram for SDSS galaxies (dots, coloured by $\log W_{\Ha}$ [\AA]) and our EL base component (white circles). All line fluxes are deredenned, but this is only relevant for the y-axis, which differs from the usual $\log \oiii/\Hb$ by $\sim -0.5$ dex. 
  }
\label{fig:BPTa}
\end{figure}

We impose that all ELs must be present simultaneously, and in proportions found in real galaxies. 
The way this idea is implemented is illustrated in Fig.\ \ref{fig:BPTa}, which shows the BPT$\alpha$ diagram, a trivial variation over the original \cite{1981BaldwinPhiTerl} diagram, where instead of \oiii/\Hb\ the y-axis shows the de-reddened  \oiii/\Ha\ flux ratio, a  physically irrelevant choice which is convenient here because we define our EL-base to have unit flux in \Ha. The dots come from a sample of over 200k SDSS galaxies analysed in \cite{Cid_2010_SDSS}, coloured by the \Ha\ equivalent width  ($W_{\Ha}$). White circles show the loci of the $n_{\rm EL} = 94$ components in our EL-base.  Their coordinates are defined as the mean log line ratios in a grid of 0.125 dex square bins containing at least 200 galaxies, plus a few extra points added to better trace the outer contours of the observed distribution. 
The values of \oii/\Ha\ and \sii/\Ha\ for each component were also defined as the average in each BPT$\alpha$ bin.

\begin{figure}
  \includegraphics[width=\columnwidth]{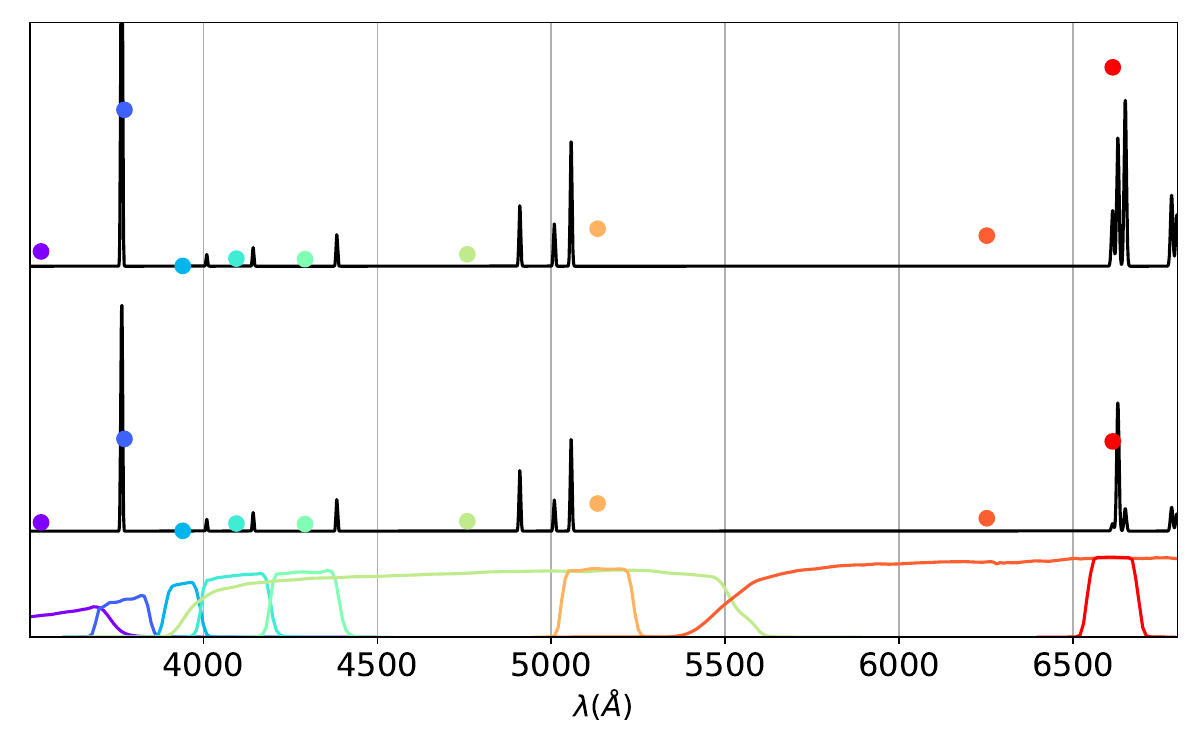}
  \caption{Example EL base spectra corresponding to a typical SF galaxy (bottom) and a retired galaxy (top) -- located at $(\log \nii/\Ha,\log \oiii/\Ha) = (-0.76,-0.26)$ and $(0.11,-0.13)$, respectively. Both are redshifted to $z = 0.01$. 
  The coloured points show the corresponding S-PLUS photometric fluxes, multiplied by 10 for clarity. The bottom curves show the S-PLUS filter transmission profiles in the range from uJAVA to J0660.
  }
\label{fig:ELbaseSpectra}
\end{figure}

The spectrum of each EL base component is built to have unitary \Ha\ flux, gaussian profiles with $\sigma = 150$ km/s, and dust-free line ratios defined as explained above. 
Nebular continuum emission is neglected. Fig.\ \ref{fig:ELbaseSpectra} shows a couple of example spectra, along with the corresponding S-PLUS photometry from uJAVA ($\lambda_{\rm pivot} = 3533$ \AA) to J0660 ($\lambda_{\rm pivot} = 6614$ \AA). The spectra are shifted to redshift $z = 0.01$, such that the main ELs fall within narrow bands, and (for this very reason) typical of the main intended applications.
As expected, ELs stand out  clearly in the narrow bands. The final effect on the photometry, of course, depends on the stellar continuum to which the lines are added. The contribution of an EL of equivalent width $W$ to a filter of width $\Delta$ is  $\approx W/(W+\Delta)$. For an SF galaxy like the bottom one in Fig.\ \ref{fig:ELbaseSpectra}, the EL contribution to J0660 (r) exceeds 50 (9)\% for $W_{\Ha} > 119$ \AA.

The top EL spectrum in Fig.\ \ref{fig:ELbaseSpectra} is typical of that of a retired galaxy, i.e., a galaxy with no ongoing star-formation nor relevant nuclear activity, and whose EL are powered only by hot, low mass, evolved stars \citep{Stasi_ska_2008}. These systems have  typically $W_{\Ha}$ of $\sim 1$ \AA\ \citep{Cid_Fernandes_2011}, in which case (and further assuming $\nii/\Ha \sim 1.3$) the EL contribution to the J0660 flux would be just $\sim 2$ percent, and negligible to the r-band.

By construction, the linear combinations of these components computed by \alstar\ approximately span the space of line ratios observed in real galaxies. The output in this case is the contribution of each component to the \Ha\ flux, to which all other lines are scaled. Mathematically, the EL model spectrum reads:  

\begin{equation}
\label{eq:M_EL_lambda}
M^{\rm EL}_\lambda = 
\sum_{j=n_\star+1}^{n_\star+n_{\rm EL}} x_j b^{\rm EL}_{\lambda,j} e^{-\tau_j q_\lambda},
\end{equation}

\noindent where $b^{\rm EL}_{\lambda,j}$ is the EL spectrum normalized to unitary $F_{\Ha}$ 
(such as those exemplified in Fig.\ \ref{fig:ELbaseSpectra})
and $x_j$ is the \Ha\ flux in component $j$. The total model combines the stellar (Eq.\ \ref{eq:M_star_lambda}) and EL (Eq.\ \ref{eq:M_EL_lambda}) components: $M_\lambda = M^\star_\lambda + M^{\rm EL}_\lambda$.

As with the stellar components, ELs are attenuated by either $\tau_{\rm ISM}(\lambda)$ or $\tau_{\rm ISM}(\lambda) + \tau_{\rm BC}(\lambda)$, depending on the configuration. 
In this study we chose to link the attenuation of the EL base components to that of young stars, which is itself tied to that of the general interstellar medium.

The fact that both ELs and $\le 10$ Myr stars undergo the same dust attenuation is the only link between stellar populations and ELs imposed in our fits.
The model can be easily modified to impose some level of consistency between these two types of components. For instance, instead of treating it as a free parameter, the \Ha\ flux can be computed in terms of the ionizing photon flux produced by the same stars used to model the continuum, effectively linking stellar populations and ELs in an astrophysical way (as done, for instance, in \citealt{BAGPIPES}). Note, however, that even this simple refinement requires full confidence on the model stellar spectra in the $< 912$ \AA\ range, as well as a recipe to deal with the dust extinction of these ionizing photons, and an assumption about their escape fraction. Given the uncertainties in all these factors, we opted not to incorporate them in this first study.

\subsubsection{Single emission line base components do not work}
\label{sec:SingleELbaseDoNotWork}

Before moving on, let us open a parenthesis to dismiss the naive idea that a compact EL base containing just one of the main ELs per component (and zero flux in the other lines) would be preferable to the scheme outlined above. 

Besides allowing for arbitrary line ratios, this smaller base would (somewhat counter-intuitively) give much more freedom to the code. Too much freedom, in fact. 
Consider, for instance, the case of a galaxy with no ELs at all (or very weak ones), but where the filter containing, say, \oiii, has a positive noise spike which prevents it from being fitted by the stellar base alone.  The code could then simply attribute this extra flux to a non-existent \oiii\ and fit the observed flux exactly, even if no other EL is found or if it implies $\oiii/\Hb > 1000$. 

This deceivingly simpler base can thus lead to nonphysical line ratios and be misused to fit (positive) noise or compensate for deficiencies in the stellar base. This was verified in numerical experiments, which also showed that even when ELs are strong the fits often lead to unrealistic line ratios.

\subsection{Equivalent width based constraints}
\label{sec:WHaN_based_constraints}

The approach of tying all ELs together and constraining them to realistic proportions should provide a reasonable estimate of their effect on the S-PLUS photometry (not only the narrow bands) and hence aid the \alstar\ fits. One should not, however, expect the ELs to be reliably retrieved, as there is not enough information in the 12  S-PLUS bands to properly constrain them in detail.

The main problem is that \Ha\ and the adjacent \nii\ lines all fall into a single filter, J0660, centered at 6614 \AA\ and with a FWHM of 147 \AA\ \citep{2019Splus}. The code should be able to handle well the combined effect of \Ha+\nii\  on the J0660 filter (as shown by \citealt{2015_Vilella_Rojo}), but disentangling them is a harder task. Tests confirmed that while \alstar\ recovers the combined  \nii6548+\Ha+\nii6584 equivalent width (hereafter \WHaNii) very accurately, the \nii/\Ha\ ratio can come out completely wrong, moving a galaxy from  left to right or vice-versa in the BPT diagram. Other lines, besides being generally weaker, do not have the ability to distinguish which wing of the BPT ``seagull'' a galaxy sits in, seriously limiting the kind of EL-based diagnostic doable with S-PLUS.

A possible strategy to deal with this limitation is to restrict the EL base to just one of the wings, chosen on the basis of ancillary data (say, X-rays). Some sort of base restriction is in order, but this is not a satisfactory general solution. Our challenge here is to mitigate this problem using only S-PLUS data.

\begin{figure*}
  \includegraphics[width=\linewidth]{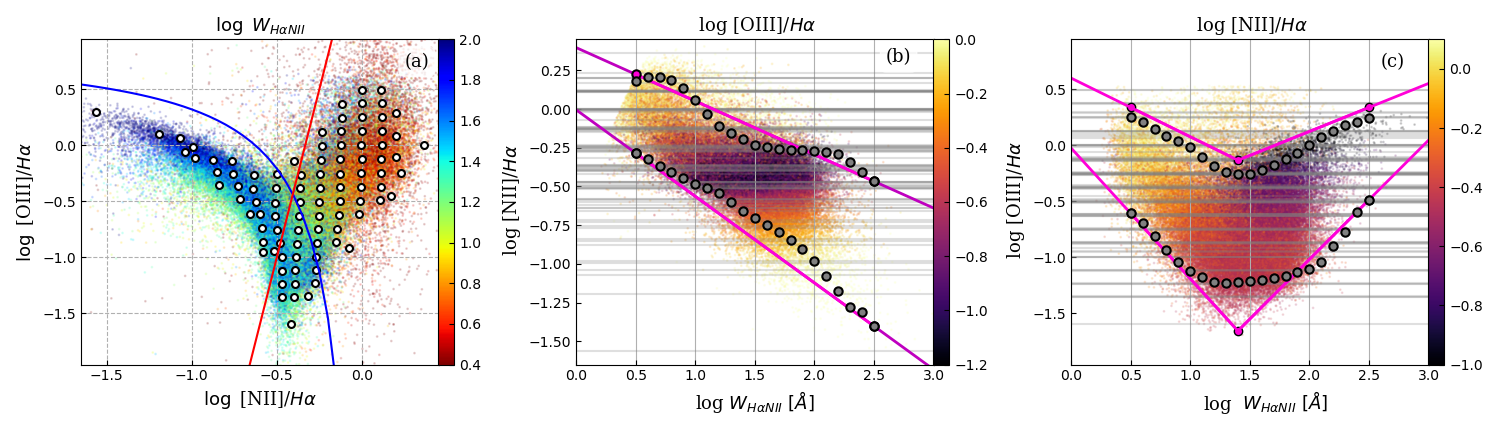}
  \caption{
  (a) As Fig.\ \ref{fig:BPTa}, but colouring by the log of $\WHaNii = W_{\Ha} + W_{\nii6548} + W_{\nii6584}$. The blue and red curves are part of the EL base restriction scheme discussed in the text (see Table \ref{tab:ELBCS}). 
   (b) $\log \nii/\Ha$ versus $\WHaNii$, coloured by $\log \oiii/\Hb$. (c) Like panel b, but for $\oiii/\Hb$, and colour-coding by $\log \nii/\Ha$. Dark circles in panels b and c track the 5 and 95 percentile curves, while horizontal gray lines mark the log-line-ratios in the full EL base. The magenta lines show the lower and upper limits imposed to limit the EL base for a given $\WHaNii$. These limits refine the left/right wing constraints shown in panel a. Together, they work as a prior which helps estimating EL properties out of the always reliable estimate of $\WHaNii$.}
\label{fig:ELBCS}
\end{figure*}

We have developed a semi-empirical method to tackle this issue. The key idea is to explore the fact that EL equivalent widths are generally larger in star-forming (SF) than in the so called AGN wing.\footnote{Despite its common use in the literature, ``AGN wing'' is a misnomer, given that the right wing in the BPT diagram also contains large proportions of galaxies where both AGN and SF activity coexist, as well as plenty of retired galaxies, which have neither ongoing SF nor an energetically relevant AGN (\citealt{Stasi_ska_2008}, \citealt{Cid_Fernandes_2011}).}
This is clearly seen in Fig.\ \ref{fig:BPTa}. The SF wing is composed almost exclusively by galaxies with $W_{\Ha} \gtrapprox 50$ \AA\ (painted in green--blue), increasing to over 100 \AA\ towards the top-left. Right wing sources have generally lower $W_{\Ha}$, including the whole population of retired galaxies (in red) with $W_{\Ha}$ of order 1 \AA. Some Seyfert 2s (towards the top-right) have \Ha\ in the range of SF galaxies, but otherwise $W_{\Ha}$ offers a good way of distinguishing left from right wing sources.

As discussed above (and demonstrated in section \ref{sec:Results_EL_properties}), we cannot trust our initial estimates of $W_{\Ha}$, but $\WHaNii \equiv W_{\Ha} + W_{\nii6548} +W_{\nii6584}$ is very reliable. We thus seek a scheme based on this robust quantity to restrict the EL base. Fig.\ \ref{fig:ELBCS}(a) repeats Fig.\ \ref{fig:BPTa},  this time colouring galaxies by \WHaNii. The general appearance of the two plots is the same. Visibly, knowledge of the value of \WHaNii\ provides valuable guidance on the whereabouts of a galaxy in the BPT diagram.

After some experimentation, we have adopted the following scheme to incorporate this prior knowledge in our analysis:

\begin{enumerate}

    \item[\em (1)] We run an initial fit with the full EL base to estimate \WHaNii. (his initial value is robust even when \nii/\Ha\ comes out wrong.

    \item[\em (2)] We first limit the EL base to the left of the blue line in Fig.\ \ref{fig:ELBCS}(a) when \WHaNii\ $> 50$ \AA, and to the right of the red one  when \WHaNii\ $< 10$ \AA. Points in the \WHaNii\ $= 10$--50 \AA\ region are not constrained by this first cut. 

    \item[\em (3)] We then compute lower and upper limits for \nii/\Ha\ as a function of \WHaNii. This was done essentially in a visual way examining Fig.\ \ref{fig:ELBCS}(b).

    \item[\em (4)] The very shape of the wings in the BPT diagram implies that limits in its x-axis translate onto the y-axis. We reinforce this implicit constraint by imposing limits on \oiii/\Ha\ as a function of \WHaNii\ (Fig.\ \ref{fig:ELBCS}(c)).
    
\end{enumerate}

The equations involved in this heuristic scheme are given in Table \ref{tab:ELBCS}. Only base components satisfying these \WHaNii-based constraints are allowed for in the fits. The most notable ``victims'' of this scheme  are Seyferts 2s, some of which are legitimate right wing sources with \WHaNii\ $> 50$ \AA\ that are forced to move to the left wing by step 2 above. More refined schemes may be able to deal with this caveat, but we chose to overlook it in this paper.\footnote{If additional (e.g. X-rays) suggest that the galaxy has an AGN then one may trivially adjust the EL base to focus on AGN-like components. 
In such cases one may further add AGN-like (say, power-law) components to the stellar population base to account for the non-stellar continuum emission of AGN, as done in the spectroscopic studies (e.g., \citealt{CidFernandes2004}).}

Finally, we note that while the specific criteria delineated here were based on integrated galaxy data from the SDSS, spatially resolved studies can straight-forwardly adjust them to better represent spaxel-based ELs, obtained from surveys like CALIFA \citep{califa} or MaNGA \citep{MANGA_BUNDY_2015}. Previous EL work on these surveys, however, suggests that our criteria would remain approximately the same for spatially resolved data (see, for instance, figure 25 in \citealt{2022_Sanchez}).

\begin{table}
\centering
\begin{tabular}{lr}
Constraint & $\WHaNii$ range \\ \hline
$\log \oiii/\Ha < 0.90 +  0.61 / (\log \nii/\Ha -0.05)$  &  $\geq50$ \AA\\
$\log \oiii/\Ha > 6 \log \nii/\Ha + 2$  &  $\leq10$ \AA        \\ \hline
$\log \nii/\Ha \leq-0.345 \min(\log \WHaNii,3) +0.397$ & any \\
$\log \nii/\Ha \geq-0.559 \max(\log \WHaNii,0) -0.005$ & any \\ \hline
$\log \oiii/\Ha \leq-0.523 \log \WHaNii + 0.598$  &  $< 25$ \AA\\
$\log \oiii/\Ha \leq 0.428 \log \WHaNii - 0.734$  &  $\geq 25$ \AA \\ 
$\log \oiii/\Ha \geq-1.164 \max (\log \WHaNii, 0) - 0.026$  &  $< 25$ \AA\\
$\log \oiii/\Ha \geq 1.063 \max(\log\WHaNii, 2.85) - 3.144$  &  $\geq 25$ \AA \\ \hline
\end{tabular}
\caption{Equations used to constrain the EL-base according to the value of $\WHaNii$ --- see also Fig.\ \ref{fig:ELBCS}.}
\label{tab:ELBCS}
\end{table}

The effective role of these criteria is to incorporate prior information on the properties of ELs in actual galaxies, hopefully aiding a more accurate retrieval of their properties out of the data offered by S-PLUS. The aim is obviously not to reach a spectroscopy-quality quantification of EL properties, but to mimic it as best as possible. Several refinements and extensions of this general idea can be explored, but we leave this for future work. Let us now put this scheme to test, first with a suite of simulations (Section \ref{sec:SDSS_simulations})
and then with some example applications to actual data (Section \ref{sec:Applications}).

\section{Simulations}
\label{sec:SDSS_simulations}

In order to test \alstar\ in the S-PLUS regime we have culled a sample of SDSS galaxies previously analysed by means of full spectral synthesis and detailed EL fitting as a reference.
We have selected 10473 SDSS galaxies out of those analysed by W19 with the \starlight\ and {\sc dobby} codes for this experiment. 
The test consists of computing the synthetic photometry of these galaxies, adding noise, running it through \alstar, and comparing its output with that obtained by W19. 

\subsection{The input}
\label{sec:TheInput}

Our sample was chosen to be uniformly distributed in $\log W_{\Ha}$, with 500 galaxies randomly drawn every 0.1 dex for $W_{\Ha}$ between 1 and 100 \AA, plus 473 galaxies above 100 \AA. It thus spans from retired galaxies, with their very weak ELs, to galaxies at the tip of the star-forming wing in the BPT diagram, where the high specific star formation rates and low gas phase metallicities lead to \Ha\ and \oiii\ in excess of  100 \AA. The choice to uniformly cover such a wide dynamic range is motivated by our desire to map the effect of the overall relevance of ELs, something that is suitably quantified by $W_{\Ha}$. Uncertainties and biases in the output ELs should increase as they become weaker (a trivial expectation that will soon be confirmed), and this sample selection scheme allows to track this effect  cleanly.

Because the SDSS spectra do not fully cover the bluest and reddest S-PLUS filters, the synthetic photometry is carried out over the models fitted by W19 (including the {\sc dobby}-based ELs), which do cover the full range.  Also, since our analysis does not incorporate GALEX fluxes, we compare our results to those obtained in the SDSS-only fits described in W19. 
All spectra are shifted to $z = 0.01$ to ensure that \Ha\ and \nii\ are both within the J0660 filter.

Finally, we note that the stellar base used by W19 is very similar to the one used here, differing only in the exact definitions of age bins.  The \starlight\ fits in that study did not account for differential extinction, however, while our \alstar\ fits do. We have verified that this does not introduce any significant change in the conclusions of this paper.
The main difference, of course, is that in W19 the ELs are masked from the spectral fits and measured a posteriori from the observed minus model residual spectra, while in our analysis stellar populations and ELs are estimated simultaneously.

\subsection{Noise}

One final ingredient we need to discuss is how noise is dealt with. We parametrize the noise amplitude by the signal-to-noise ratio in the r-band ($SN_r$). Errors in the other bands are scaled to that in the r-band according to $\epsilon_\lambda/\epsilon_r =$ 3.37, 4.77, 7.44, 6.41, 5.70, 1.54, 3.27, 1.03, 0.99, 1.80, 1.33 for the uJAVA, J0378, J0395, J0410, J0430, g, J0515, J0660, i, J0861, and z bands, respectively. This error spectrum is derived from statistics of thousands of galaxies observed by S-PLUS, and is similar to the one obtained from spaxel-based statistics in the S-PLUS datacubes analysed so far, so we take it as representative of the survey as a whole. Note that, unsurprisingly, the blue bands are much noisier than the red ones. 
On average over our test sample, the five bluest bands, which contain \oii\ and the age-sensitive 4000 \AA\ break, have typical signal-to-noise ratios $5.6 \times$ smaller than in the r-band.

Three values of $SN_r$ are used in our simulations: 25, 50, and 100. Our main intended applications are for galaxies of $z$ low enough so that \nii6584 is still within the J0660 filter, which corresponds to $z < 0.018$. 
The discussion focuses on results obtained for $SN_r = 50$, which is a compromise between the $SN_r$ of spatially integrated data and that attainable for spatially resolved fluxes in this redshift range. 

Each galaxy is fitted $n_{\rm MC} = 100$ times, perturbing the input photometry with gaussian noise with amplitude defined by $\epsilon_\lambda$ scaled to reach the target $SN_r$. The intended role of these MC runs is to estimate the uncertainty in any given output property (say, the stellar mass $M_\star$). In these simulations, however, we also use them to estimate the property itself, which we do using the median value over the MC runs.
The reason why this is preferable to using the best-fit result is that the input data is built out of models which (except for the ELs) use essentially the same ingredients used in the \alstar\ fits. Despite the huge difference between the $\sim 4000$ spectroscopic fluxes used in the original full spectral analysis and the 12 S-PLUS bands used here, it is more appropriate to report the results obtained from the analysis of the noise-perturbed runs than those resulting from fits of idealised noiseless data.

Let us now see what these simulations teach us.

\subsection{Stellar population properties}

\begin{figure*}
  \includegraphics[width=\linewidth]{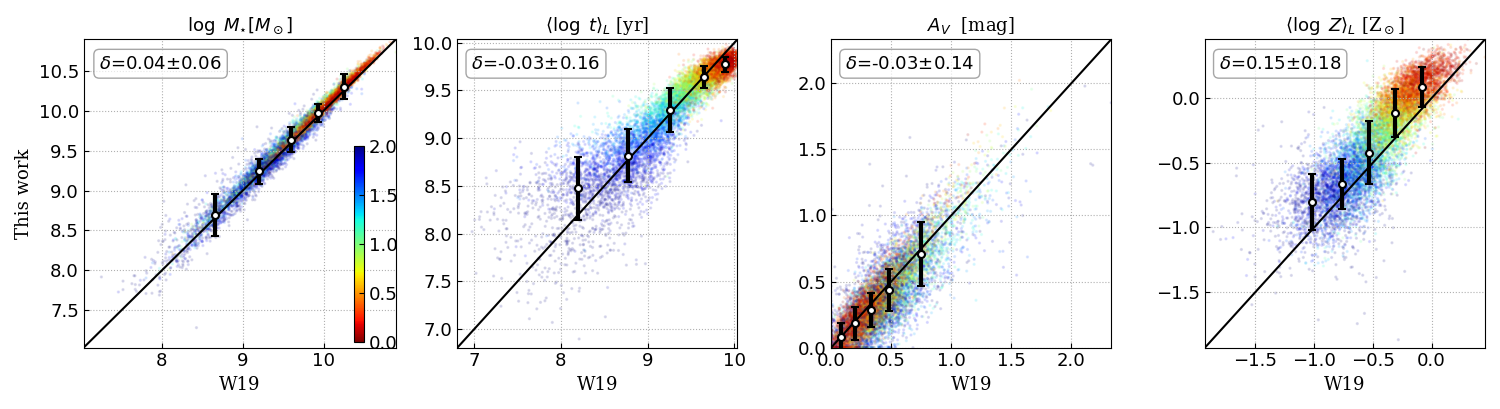}
  \caption{Comparison of the stellar population properties derived by W19 on the basis of full spectral fitting (x-axis) and those obtained here with S-PLUS photometry and \alstar\ (y-axis) for our $\sim 10$k SDSS-based test sample. Points are colour-coded according to the input value of $\log W_{\Ha}$ (inset colorbar in the left panel). The y-axis represents the median values obtained in 100 MC runs for data perturbed with $SN_r = 50$. 
  Error bars illustrate the typical (median $\sigma_{\rm NMAD}$) dispersion between output and input values for bins along the x-axis.
  As in W19, the luminosity weighting in $\langle \log t \rangle_L$ and $\langle \log Z \rangle_L$ is made with the (rest-frame) flux around 5635 \AA. 
}
\label{fig:Simulations_Stellar}
\end{figure*}

\begin{table}
\centering
\begin{tabular}{lrrr}
Property                                &   $SN_r = 25$   &     $SN_r = 50$  &        $SN_r = 100$ \\ \hline
$\log M_\star [M_\odot]$                &  $ 0.03\pm 0.06$  & $ 0.04\pm 0.06$ & $ 0.04\pm 0.05$ \\ 
$\langle \log t \rangle_L [{\rm yr}]$   &  $-0.08\pm 0.18$  & $-0.03\pm 0.16$ & $-0.01\pm 0.15$  \\
$\langle \log t \rangle_M [{\rm yr}]$   &  $0.07 \pm0.10$   & $0.08\pm 0.10$ &  $0.07 \pm0.09$ \\
$\langle \log Z \rangle_L  [Z_\odot]$   &  $0.26 \pm0.19$   & $0.15\pm 0.18$ &  $0.09 \pm0.17$ \\
$\langle \log Z \rangle_M [ Z_\odot]$   &  $0.31 \pm0.26$   & $0.19\pm 0.20$ &  $0.11 \pm0.17$ \\
$A_V [\rm mag]$                         &  $-0.01\pm 0.19$  & $0.03\pm 0.14$ &  $-0.02\pm 0.11$  \\ \hline
\end{tabular}
\caption{Statistics of the $\delta =$ output minus input for different properties and $S/N$ ratios in the r-band. The entries list median $\pm \sigma_{\rm NMAD}$ of $\delta$ over the 10473 galaxies in the test sample.}
\label{tab:Simulations_StPopsstats}
\end{table}

Despite our particular interest in ELs, we also want to test how well the code retrieves basic stellar population properties like the stellar mass, light and mass weighted mean ages and metallicities, and dust content, so let us check these first.

Fig.\ \ref{fig:Simulations_Stellar} compares stellar masses ($M_\star$), luminosity weighted mean log stellar age ($\langle \log t \rangle_L$) and metallicity ($\langle \log Z \rangle_L$), and stellar extinction ($A_V$) obtained by W19 with those derived here. The \alstar-values (in the y-axis) are the median over the 100 MC fits for each galaxy in the $SN_r = 50$ runs. 
Points are coloured according to the input $\log W_{\Ha}$ to map the overall influence of ELs on the photometry (and hence on the \alstar\ fits).
Each panel lists the sample median $\pm$ the normalized median absolute deviation ($\sigma_{\rm NMAD}$)\footnote{$\sigma_{\rm NMAD}(x) = 1.4826 \times {\rm median}(|x - {\rm median}(x)|)$, a robust descriptor of the width of a distribution, equivalent to the standard deviation for Gaussian data.}
of $\delta$, where $\delta$ denotes the output minus input difference in the corresponding property. 
The error bars break down the statistics of $\delta$ for bins along the $x$-axis.

The figure shows a satisfactory level of agreement, especially  considering that \alstar\ used only the 12 S-PLUS bands and had the extra burden of simultaneously accounting for ELs, whereas W19 values come from full spectral fits where ELs were  masked.

As expected from previous experiments of this sort (e.g., \citealt{2003Bell}, \citealt{Cid2005}, \citealt{2011Taylor}), the stellar mass is very well recovered (left panel), with a difference between output and input $\log M_\star$ of just $\delta = 0.04 \pm 0.06$ dex over the whole sample. The plot also shows that sources with strong ELs (blue points) are more dispersed around the one-to-one line. Quantitatively, the scatter ($\sigma_{\rm NMAD}$) is 0.10 dex for galaxies with $W_{\Ha} > 30$ \AA,  and 0.04 dex for those with weaker \Ha. This difference is partly due to the effects of ELs on the fits, and partly a consequence of the fact that the star-formation histories of galaxies become increasingly skewed towards younger ages as $M_\star$ decreases (the so called ``downsizing'' phenomenon -- e.g., \citealt{2004Heavens}). This broader mixture of populations of different ages naturally induces larger uncertainties in the mass-to-light ratio, and hence on $M_\star$.

The values of $\langle \log t \rangle_L$ (the ``mean age'', for short --- see, e.g., GD21 for its definition), agree reasonably well, with a bias of just -0.03 and a scatter of 0.16 dex. Different factors contribute to this scatter: differences in the age bins and differential-extinction set-ups used in W19 and here, the huge compression in input information (full spectrum vs.\ 12 bands), the need to account for ELs as well as stellar populations simultaneously, and noise. This last factor entails a subtlety. As already noted, though the signal-to-noise in the r-band is 50, the blue filters are much noisier. F0395 and F0410, in particular, have errors $\sim 7$ times larger than in r. These two bands trace the 4000 \AA\ break, the most powerful stellar population age indicator in the optical range (e.g., \citealt{1983Bruzual_earlytype}, \citealt{2003KalffmannSDSS}), inevitably affecting the estimates of $\langle \log t \rangle_L$.

The third panel in Fig.\ \ref{fig:Simulations_Stellar} compares the stellar extinction $A_V$ values of W19 to the ones found here ($A_V = \tau_{\rm ISM} \times 2.5 \log e$). The agreement is good: $\delta A_V = -0.03 \pm 0.14$ mag. Galaxies with strong ELs tend to scatter more around the one-to-one line ($\delta A_V = -0.04 \pm 0.20$ mag for $W_{\Ha} > 30$ \AA), but even for these the differences are acceptable given the disparity in the amount of information used in the $x$ and $y$ axes.

The agreement in (luminosity weighted mean log) metallicities (right panel in Fig.\ \ref{fig:Simulations_Stellar}) is better than anticipated for photometric data, with a scatter of just 0.18 dex. 
The offset of $+0.15$ dex originates from a tendency of the MC runs to produce a broad and positively skewed distribution of $\langle \log Z \rangle_L$ values. In any case, galaxy mean stellar metallicities are hard to estimate even with spectroscopy, so the level of agreement found here is more than acceptable.

The MC runs in \alstar\ provide a measure of the uncertainties in individual galaxy properties. These MC-based uncertainties are of the same order as the dispersion in output minus input $\delta$ values reported above, but with a tendency to be larger (by factors of $\sim 1.3$--2).
These differences between MC and empirical (i.e., $\delta$-based) estimates of uncertainties decrease as $SN_r$ increases. Taking the latter as a fiducial reference, the general conclusion here is that \alstar\ tends to overestimate the uncertainties in its derived properties, but not by a great margin. 

The statistics for the $SN_r = 25$, 50, and 100 simulations are given in Table \ref{tab:Simulations_StPopsstats}. As expected, the differences increase for noisier data and vice-versa. For most properties the changes in  $\delta$ are small, such that results are satisfactory even for $SN_r = 25$. The property whose dispersion varies the most is $A_V$, which nearly doubles from $SN_r = 100$ to 25. The table also shows that mass-weighted mean ages and metallicities are less robust than luminosity weighted ones. This happens even in full spectral fits (see table 1 of \citealt{Cid2005} for an example), and is ultimately a consequence of the highly non-linear mass-luminosity relation of stars.

From the previous work reported in GD21 we already knew that \alstar\ has a good performance in extracting stellar population properties out of J-PAS data. The results above show that it also performs well under the S-PLUS regime, and without resorting to masking ELs.

\subsection{EL properties}
\label{sec:Results_EL_properties}

Let us now move to the main part of this study. We start the comparison of input and output EL properties by proving that \WHaNii\ is indeed very well recovered by \alstar, a premise of the whole scheme described in \ref{sec:WHaN_based_constraints}. We then investigate how reliably  
individual ELs are retrieved, and how key line ratios are recovered with our methodology.

\subsubsection{$\Ha + \nii$}

\begin{figure}
  \includegraphics[width=\columnwidth]{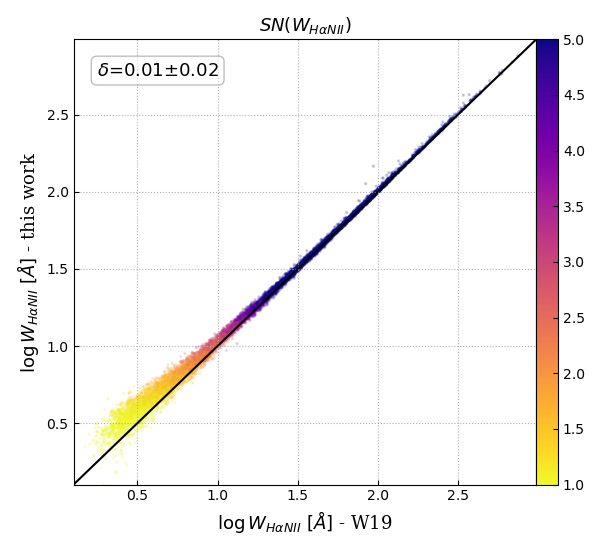}
  \caption{Comparison of the input (SDSS + \starlight\ + {\sc dobby}) and  output (S-PLUS + \alstar) values of $\WHaNii$ (the combined equivalent width of \Ha\ and \nii6548,6584) for the $SN_r = 50$ simulations. 
  Points are coloured by the \alstar-based estimate of the $S/N$ of the combined \Ha\ and \nii6548, 6584 equivalent widths.
}
\label{fig:Simulations_ELs_WHaN}
\end{figure}

Fig.\ \ref{fig:Simulations_ELs_WHaN} plots the input W19 values of \WHaNii\ against our results for the simulations with $SN_r = 50$. The colouring scheme traces the $S/N$-ratio of \WHaNii, as inferred from the median (the signal) and $\sigma_{\rm NMAD}$ (the noise) values over the MC runs for each galaxy. 

The agreement between SDSS and S-PLUS-based measurements is excellent, extending down to \WHaNii\ values in  the retired galaxy regime ($\WHaNii \lessapprox 5$ \AA).\footnote{\cite{Cid_Fernandes_2011} define retired galaxies as those with $W_{\Ha} < 3$ \AA. Given that $\nii/\Ha \sim 1$--2 in these systems, and that the peak in their $W_{\Ha}$ distribution is at $\sim 1$ \AA, it is reasonable to classify $\WHaNii \lessapprox 5$--6 \AA\ systems as retired. } The statistics of the output minus input difference is $\delta \log \WHaNii = 0.01 \pm 0.02$ dex for the 10k galaxies. The bias and scatter increase as \WHaNii\ decreases, but remain small ($0.08 \pm 0.06$ dex) even when the input $\WHaNii < 5$ \AA. 
The results are also excellent for lower $SN_r$ (see Table \ref{tab:Simulations_ELStats}).
This confirms the expectation that, because they are all sampled in the J0660 filter (for the redshifts considered in this study), the combined flux of \Ha\ and \nii\ lines should be well recovered.

Careful inspection of the colours in Fig.\ \ref{fig:Simulations_ELs_WHaN} suggests that the \alstar\ MC-based $S/N$ of $\WHaNii$ seem somewhat low for the level of agreement between input and output seen in the plot. This over-estimation of the uncertainty is analogous to that identified in the examination of stellar population properties.

\begin{figure*}
  \includegraphics[width=2\columnwidth]{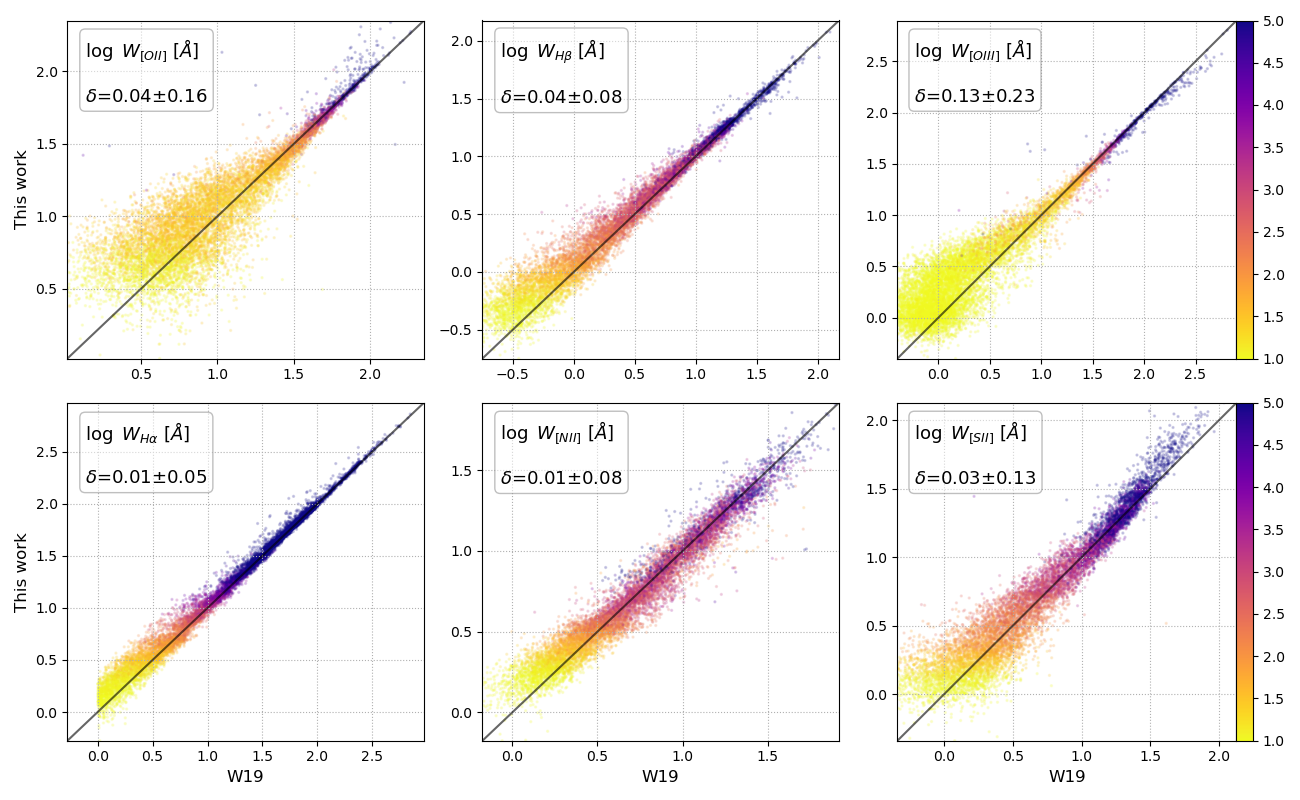}
  \caption{Comparison of input and output EL log equivalent widths for the $SN_r = 50$ simulations.
  As in Fig.\ \ref{fig:Simulations_ELs_WHaN}, the colour scale reflects the median/$\sigma_{\rm NMAD}$ ratio of the corresponding line.
  }
\label{fig:Simulations_ELs_Ws}
\end{figure*}

\subsubsection{Individual lines}

Fig.\ \ref{fig:Simulations_ELs_Ws} shows how individual ELs are recovered, including \Ha\ and \nii\ separately. Unlike for \nii+\Ha, we now see substantial differences between input and output, but mainly in the bottom-left of the plots, where lines are intrinsically weak. In all cases the $\sigma_{\rm NMAD}$ of the $\delta =$ output minus input is better than 0.23 dex for the whole sample, improving to $ < 0.15$ dex when only lines stronger than 5 \AA\ are considered (Table \ref{tab:Simulations_ELStats}).

The best results are obtained for \Ha, \Hb\ and \nii, with $\sigma_{\rm NMAD} = 0.05$, 0.08 and 0.08 dex, respectively. Despite being the second strongest EL on average in our 10k sample (with $\overline{W}_{\oii} = 16$ \AA), \oii\ is not so well recovered ($\delta = 0.04 \pm 0.16$ dex). As with the mean ages discussed above, this is due to the much larger photometric errors of the blue filters ($\sim 5 \times$ larger in J0378, the filter containing \oii, than in r). The statistics barely change restricting the sample to $W_{\oii} > 5$ \AA\ sources (Table \ref{tab:Simulations_ELStats}), but for \oii\ stronger than 10 \AA\ the agreement improves to $\delta \log W_{\oii} = 0.02 \pm 0.09$ dex. As intuitively expected, stronger lines are better retrieved.

The worst results are found for \oiii\ ($\delta = 0.13 \pm 0.23$ dex), but, again, mainly when it is intrinsically weak. For galaxies where $W_{\oiii} > 5$ \AA\ we find a perfectly acceptable match: $\delta \log W_{\oiii} = 0.00 \pm 0.09$ dex. Remarkably, even \sii\ is well recovered, despite not falling in a narrow band, and thus having a small effect on the photometry. The reason we recover it so well is not because of its photometric relevance (typically $< 1\%$ of the r-band flux), but because our base ties it to other ELs in an empirically based way.

The colour scale in Fig.\ \ref{fig:Simulations_ELs_Ws} reflects the MC-based  $S/N$ ratio of the corresponding EL. As for stellar population properties and $\WHaNii$, the MC-uncertainties in the individual ELs are somewhat over-estimated with respect to the  empirical dispersion in the output vs.\ input values.

Results for the $SN_r = 25$, 50, and 100 simulations are given in Table \ref{tab:Simulations_ELStats}. The right-most columns of the table re-evaluate the statistics of $\delta \log W$ considering only lines detected with $W > 5$ \AA. The numbers confirm the visual impression from Fig.\ \ref{fig:Simulations_ELs_Ws} that ELs above this threshold can be considered very reliable.

Inspecting  Table \ref{tab:Simulations_ELStats} one realises that (except for $W_{\nii}$ when restricted to $> 5$ \AA) all $W$'s are positively biased, even if by negligible margins in most cases. This tendency to over-predict the strength of ELs is inherent to \alstar. By construction, ELs can only contribute positively to the photometric fluxes, and thus may be used by the code to compensate for positive noise fluctuations if that helps improving the fit (see Section \ref{sec:SingleELbaseDoNotWork}). 
Note, however, that this effect is only relevant when ELs are intrinsically weak, and thus hard to be accurately retrieved.

\begin{table*}
\centering
\begin{tabular}{lrrrrrrrrr}
                           &            \multicolumn{3}{c}{All galaxies}            &             \multicolumn{3}{c}{$W \geq 5$ \AA}            \\ 
Property                   &  $SN_r = 25$~     &      50 ~~~~~~~ &       100 ~~~~~~~&      25 ~~~~~~~  &     50 ~~~~~~~     & 100 ~~~~~~~ \\ \hline
$\log \WHaNii$      &  $ 0.03\pm 0.04$  & $ 0.01\pm 0.02$  &  $ 0.01\pm 0.01$  &  $ 0.03 \pm 0.03$  & $ 0.01\pm 0.02$  &  $ 0.00\pm 0.01$ \\
$\log W_{\oii}$            &  $ 0.03\pm 0.19$  & $ 0.04\pm 0.16$  &  $ 0.03\pm 0.14$  &  $ 0.05 \pm 0.16$  & $ 0.05\pm 0.15$  &  $ 0.03\pm 0.13$ \\
$\log W_{\Hb}$             &  $ 0.04\pm 0.09$  & $ 0.04\pm 0.08$  &  $ 0.03\pm 0.07$  &  $ 0.03 \pm 0.04$  & $ 0.02\pm 0.04$  &  $ 0.00 \pm0.03 $\\
$\log W_{\oiii}$           &  $ 0.13\pm 0.26$  & $ 0.13\pm 0.23$  &  $ 0.11\pm 0.20$  &  $ 0.06 \pm 0.18$  & $ 0.00\pm 0.09$  &  $ 0.00\pm 0.05$  \\
$\log W_{\Ha}$             &  $ 0.04\pm 0.06$  & $ 0.01\pm 0.05$  &  $ 0.0 \pm 0.04$  &  $ 0.02 \pm 0.04$  & $ 0.01\pm 0.03$  &  $ 0.00 \pm0.03 $ \\
$\log W_{\nii}$            &  $ 0.02\pm 0.11$  & $ 0.01\pm 0.08$  &  $ 0.01\pm 0.07$  &  $ -0.03 \pm 0.08$  & $ -0.01\pm 0.07$ &  $  0.01\pm 0.06$  \\
$\log W_{\sii}$            &  $ 0.06\pm 0.14$  & $ 0.03\pm 0.13$  &  $ 0.02\pm 0.12$  &  $ 0.03\pm 0.10$  & $ 0.02\pm 0.09$  & $ 0.01\pm 0.08$ \\
$\log \nii/\Ha$            &  $-0.03\pm 0.13$  & $-0.01\pm 0.12$  &  $0.00 \pm 0.12$  &  $-0.05 \pm 0.11$  & $-0.02\pm 0.10$  &  $0.01 \pm0.09 $ \\
$\log \oiii/\Hb$           &  $0.06\pm 0.26$  & $0.07\pm 0.23$  &  $0.05 \pm 0.19$  &  $0.03 \pm 0.16$  & $0.00\pm 0.09$ &  $0.00\pm 0.06$  \\
$\log \Ha/\Hb$             &  $-0.01\pm 0.06$  & $-0.02\pm 0.06$  &  $-0.03\pm 0.06$  &  $-0.02 \pm 0.04$  & $-0.01\pm 0.04$  &  $-0.01\pm 0.04$   \\
$\log \oiii/\oii$          &  $0.09\pm 0.23$  & $0.07\pm 0.20$  &  $0.05 \pm 0.19$  &  $0.04 \pm 0.16$  & $0.01\pm 0.09$ &  $ 0.01 \pm0.06 $  \\ 
O3N2                  &  $0.09\pm0.30$    & $0.09\pm0.26$   &  $0.08\pm0.23$       &$0.08\pm0.26$    &$0.00\pm0.17$     &$-0.02\pm0.13$      \\ \hline
\end{tabular}
\caption{Statistics of the $\delta =$ output minus input for EL properties and different $S/N$ ratios in the r-band. As in Table \ref{tab:Simulations_StPopsstats}, each entry gives the median $\pm$ $\sigma_{\rm NMAD}$ of $\delta$ for different properties (all in dex). For columns 2, 3 and 4 the statistics is performed with all 10k galaxies in the test sample, while in the last three columns only sources where the corresponding line is detected with $W > 5$ \AA\ are included. For the last 5 rows, the $> 5$ \AA\  limit is applied to all ELs involved in the ratio.
}
\label{tab:Simulations_ELStats}
\end{table*}

\subsubsection{Line ratios}
\label{sec:LineRatios}

\begin{figure}
  \includegraphics[width=\columnwidth]{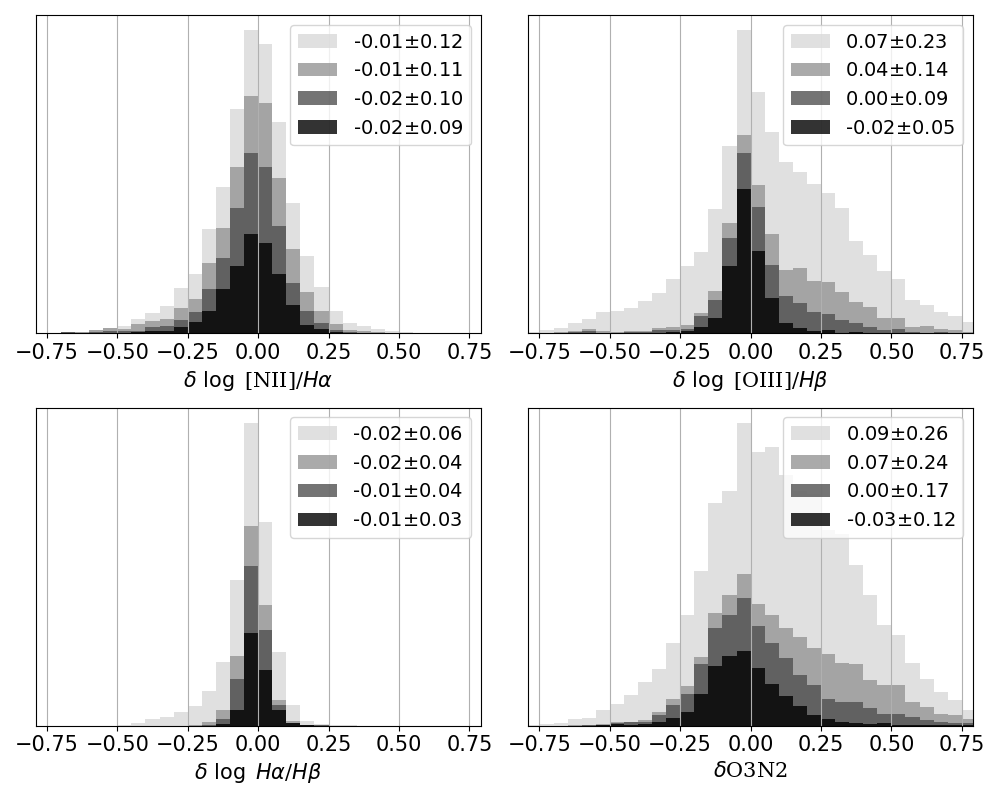}
  \caption{Histograms of the output minus input log line ratios for $SN_r = 50$ simulations. The lighter histograms are for the 10k galaxies. The progressively darker ones are for sub-samples defined by imposing $W> 3$, 5, and 10 \AA\ to the lines involved in the panel. The values in the legends are the median $\pm\, \sigma_{\rm NMAD}$. As expected, the ratios become progressively more reliable as the lines get stronger. 
  }
\label{fig:Simulations_ELs_Ratios}
\end{figure}

Finally, let us investigate some flux ratios recovered by \alstar. Estimating ratios is always more delicate than estimating the numerator and denominator separately, and are intrinsically biased (\citealt{1994Rola_Pelat}, \citealt{2016Wesson}). Still, because of their astrophysical relevance, it is important to test our ability to estimate line ratios out of S-PLUS data. 

Fig.\ \ref{fig:Simulations_ELs_Ratios} shows histograms of the output minus input    \nii/\Ha, \oiii/\Hb, and \Ha/\Hb\ log ratios, as well as for ${\rm O3N2} = \log \{ (\oiii/\Hb) / ( \nii/\Ha) \}$, a popular nebular metallicity indicator \citep{2004Pettini}. The light shaded distributions are for the full 10k sample, while the darker ones are for sub-samples constructed by requiring that the \alstar-based $W$'s are $> 3$, 5, and 10 \AA\ (light gray, dark gray, and black, respectively) for all lines involved. The statistics (median $\pm \sigma_{\rm NMAD}$) of $\delta$ are given in each panel (see also Table \ref{tab:Simulations_ELStats}).

The figure shows that \nii/\Ha\ is well retrieved, with negligible bias and $\sigma_{\rm NMAD}$ decreasing from 0.12 dex for the whole sample to 0.09 dex when both \nii\ and \Ha\ have $W > 10$ \AA. This success is not surprising, given the robustness of $\WHaNii$ and the correlation between \nii/\Ha\  and $\WHaNii$ on which we based our empirical EL priors. \Ha/\Hb\ is also very well recovered, despite the fact that \Hb\ is only covered by a broad band (g). This is only possible because \Hb\ is directly tied to \Ha\ in our analysis.

Ratios involving \oiii\ are more problematic, as illustrated by the $+0.07$ bias and $\pm 0.23$ dex dispersion in $\delta \log \oiii/\Hb$ for the full sample. This happens because our limits on \oiii/\Ha\ are not as constraining as those for \nii/\Ha\ (see Fig.\ \ref{fig:ELBCS}). Also, \oiii\ is alone in filter J0515, which makes it prone to be misused to fit positive noise. Again, most of these caveats apply to the weak line regime. As illustrated by the darker histograms in Fig.\ \ref{fig:Simulations_ELs_Ratios}, imposing a $W > 5$ \AA\ cut in \oiii\ and \Hb\ eliminates the bias and reduces the scatter to just 0.09 dex.
O3N2 is the most uncertain index discussed here. Only when all four lines involved are stronger than 10 \AA\ its scatter approaches 0.1 dex, though it may still be of statistical value for galaxies with weaker lines ELs.

We thus find that relatively reliable line ratios are obtainable with this method, particularly when focusing on sources with strong lines ($W > 5$ \AA). This offers the prospect of estimating nebular metallicities (with either \nii/\Ha\ or O3N2) as well as EL dust reddening (via \Ha/\Hb).

\subsection{Degeneracies}
\label{sec:Degeneracies}

\begin{figure*}
    \centering
    \includegraphics[width=\textwidth]{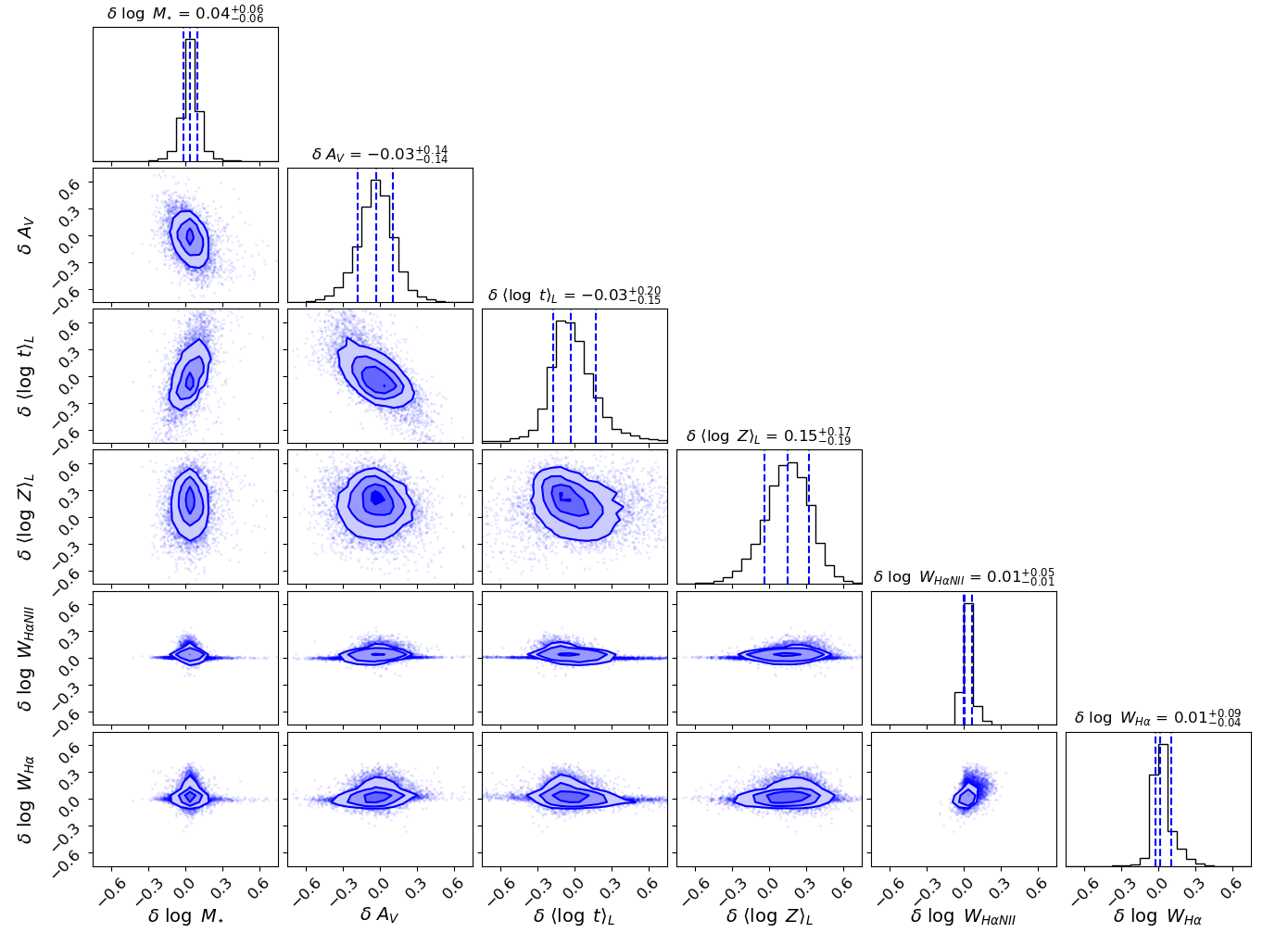}
    \caption{Corner plot \citep{corner} illustrating the degeneracies between different stellar and EL properties for the $SN_r = 50$ simulations described in the text. Each panel show the $\delta = $ output $-$ input difference between our estimate of a given property (the output) and that by W19 (the input) against the $\delta$ in another property.  The four contour levels correspond to 0.5, 1, 1.5, and $2 \sigma$-equivalent levels. Plots along the diagonal show the histograms of the $\delta$'s of the property in the x-axis, with dashed blue lines marking the 16, 50, and 84 percentiles.
     }
    \label{fig:corner_plot}
\end{figure*}

Spectral synthesis analysis is notoriously prone to degeneracies between age, dust and metallicity (see \citealt{Conroy2013} and references therein). Because of the disparity in the amount of input information, these degeneracies are more severe for photometric than for full spectral synthesis. The simulations above show that our method is able to recover spectroscopic-based estimates of stellar and EL properties to a good level of accuracy. It is nevertheless useful to inspect degeneracies among our estimated properties, as this has not been done before for S-PLUS and the methodology employed  here.

Fig.\ \ref{fig:corner_plot} addresses this issue. The off diagonal panels in this corner plot show how the difference ($\delta$) between our estimate and the W19 one for a given property correlates with the $\delta$ in another property. The panels involving $\log M_\star$,  $\langle \log t \rangle_L$, $\langle \log Z \rangle_L$, and $A_V$ show well known degeneracies inherent to stellar population properties. For instance, $\delta \langle \log t \rangle_L$ anti-correlates with both $\delta A_V$ (the age-dust degeneracy) and $\delta \langle \log Z \rangle_L$ (the age-metallicity degeneracy), reflecting the fact increasing values of any one of these properties leads to a redder spectrum, so that over-estimated ages are compensated by under-estimated extinction or metallicity. Similarly, the positive correlation between $\delta \langle \log t \rangle_L$ and $\delta \log M_\star$ is due to the well established fact that older populations have larger mass-to-light ratios.

The main novelty in our methodology is the simultaneous fitting of both stellar and EL properties, an approach which could, in principle, spur new modes of degeneracies. The panels involving  $\log W_{\Ha}$ and $\log \WHaNii$ show that this is {\em not} the case. Neither $\delta \log W_{\Ha}$ nor $\delta \log \WHaNii$ correlate with any of the stellar population properties, and the same is true for other ELs. 
We attribute this to the fact that the EL base spectra, with their ups and downs from one filter to the next (Fig.\ \ref{fig:ELbaseSpectra}), are completely unlike the stellar spectra, which change more smoothly with $\lambda$. This dissimilarity prevents confusion between EL and stellar components, explaining the lack of correlation seen in Fig.\ \ref{fig:corner_plot}.
We thus conclude that, at least in the empirically motivated way explored in this study, fitting for both stellar EL properties does not lead to significant degeneracies other than those typical of stellar population work.

\section{Example applications}
\label{sec:Applications}

The performance of \alstar\ in the simulations of the previous section confirmed our basic expectations with respect to the stellar population properties, and exceeded them with respect to ELs. In this section we present some example applications to actual S-PLUS data. The goal is merely to illustrate the quality of spectral fits obtained and the reasonability of the inferred properties. 

For the applications below, a couple of modifications are made with respect to the \alstar\ runs in  Section \ref{sec:SDSS_simulations}. First, the smallest metallicity is $0.2 Z_\odot$, reducing the base from 7 to 5 $Z$ values. The lowest $Z$ models are very rarely used anyway (see Fig.\ \ref{fig:Simulations_Stellar}). 
A second difference is that, for each band, we modify the input photometric errors $\epsilon_\lambda$ by adding (in quadrature) a minimum error to ensure that the signal-to-noise ratio does not exceed $SN_{\rm max} = 100$. This is equivalent to accounting for an uncertainty in the models, i.e., to acknowledge that the method as well as its ingredients are not perfect enough to fit perfect data. This affects mostly the MC statistics and only in the limit of high SN.

\subsection{NGC 1365 and NGC 1379} \label{sec:3Galaxies}

\begin{figure}
  \includegraphics[width=\columnwidth]{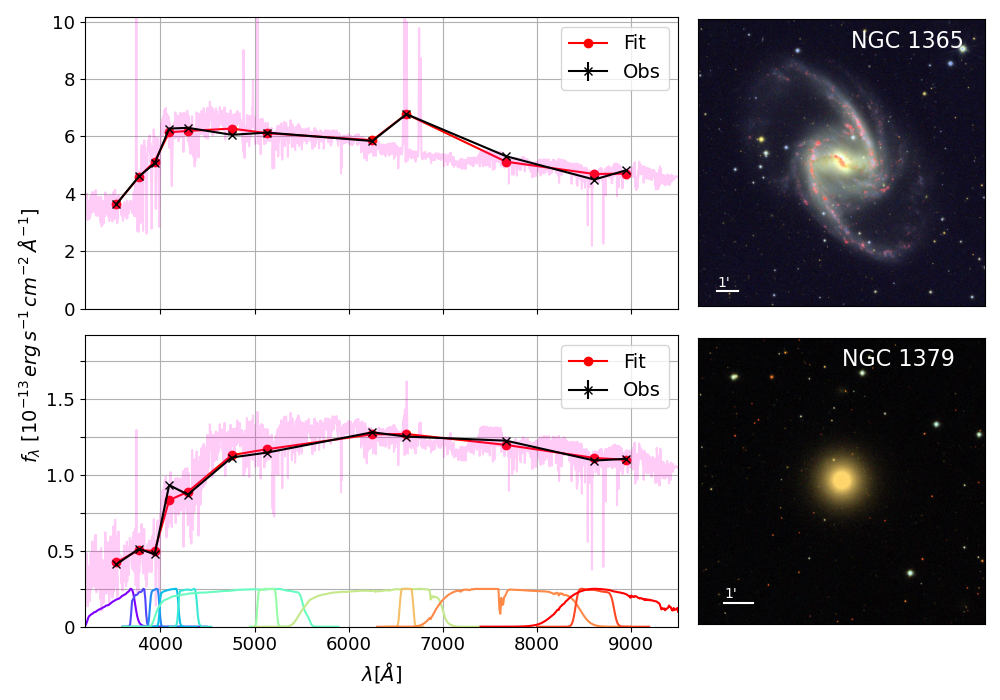}
  \caption{Example \alstar\ fits of S-PLUS data for two galaxies. The data is plotted in black crosses, while the model photometric  fluxes are plotted as red circles. The model spectrum is plotted in magenta. The images are composites built with the J0660, r, and g fluxes for NGC 1365, and i, r and g for NGC 1379, in the R, G, and B channels, respectively.
  }
\label{fig:3Galaxies}
\end{figure}

We have chosen  NGC 1379 (an elliptical galaxy) and NGC 1365 (a barred-spiral), both in the Fornax cluster, to illustrate the application of \alstar\ to integrated light S-PLUS data. Details on these and other Fornax integrated photometry, based on the S-PLUS Data Release 4 (Herpich et al in prep), will be presented in a forthcoming article (Smith Castelli et.\ al., in prep). The observed photometry is shown in black in Fig.\ \ref{fig:3Galaxies}, along with the \alstar\ fits\ (in red) and composite images. 

The fits to the photometry are visibly excellent, with a mean absolute deviation between data and model of 1.7\% and 2.8\% for NGC 1379 and NGC 1365, respectively. 
Our fits yield $\log M_\star/M_\odot = 10.48$ for NGC 1379 and 10.86 for NGC 1365. These values compare very well with previous estimates. For instance, \cite{2019IodiceVST} derived a value of 10.42 for NGC 1379 on the basis of VST photometry and the \cite{2011Taylor} recipe for $M/L_i$ as a function of $g - i$. Similarly, \cite{2019Raj} and \cite{2022ASu} estimate $\log M_\star/M_\odot = 10.81$ and 11.00 for NGC 1365.

A detailed comparison with spectroscopy-based stellar and EL properties is complicated by the fact that, because of their proximity, such studies cover only part of the galaxy, whereas the data in Fig.\ \ref{fig:3Galaxies} correspond to the whole galaxies. Let us nevertheless make a rough comparison with the MUSE-based results for NGC 1379 presented in \cite{2019IodiceMuse}. Their stellar population analysis indicates a $\log$ age/yr of 10.11 over the inner $0.5R_e$  ($\sim 14.3$ arcsec), whereas our whole-galaxy estimate suggest similar ages: $\langle \log t \rangle_L = 10.01$ and $\langle \log t \rangle_M = 10.09$ yr. Considering the very different methodologies, very different data, and  different apertures, these estimates can be seen as broadly compatible, in the sense that both point to a dominance of old stellar populations. Regarding metallicities, our whole-galaxy estimates are $\langle \log Z/Z_\odot \rangle_L = -0.12$ and $\langle \log Z/Z_\odot \rangle_M = -0.17$, while \cite{2019IodiceMuse} obtain $-0.13$ over the central $0.5R_e$.

Regarding ELs, the \alstar\ fits of NGC 1379 yield $\WHaNii = 5.8$ \AA, $W_{\Ha} = 2.5$ \AA, $\nii/\Ha = 0.98$ for the whole galaxy. These values suggest a retired galaxy classification \citep{Cid_Fernandes_2011} and  are consistent with what would be expected from an early-type galaxy. For NGC 1365 we find integrated EL properties consistent with a star-forming galaxy, with 
$\WHaNii = 40.4$ \AA, $W_{\Ha} = 31.02$ \AA, $\nii/\Ha = 0.22$, $\oiii/\Hb = 0.79$. As with NGC 1379, previous spectroscopic work on this galaxy focuses on its inner regions, hindering a proper comparison with these galaxy-wide estimates. 

These two examples fulfil the goal of illustrating the performance of \alstar\ with real S-PLUS integrated-light data. Let us now examine the potential of S-PLUS + \alstar\ for spatially resolved work.

\subsection{Datacubes}
\label{sec:DataCubes}

\begin{figure*}
    \centering
    \includegraphics[width=\textwidth]{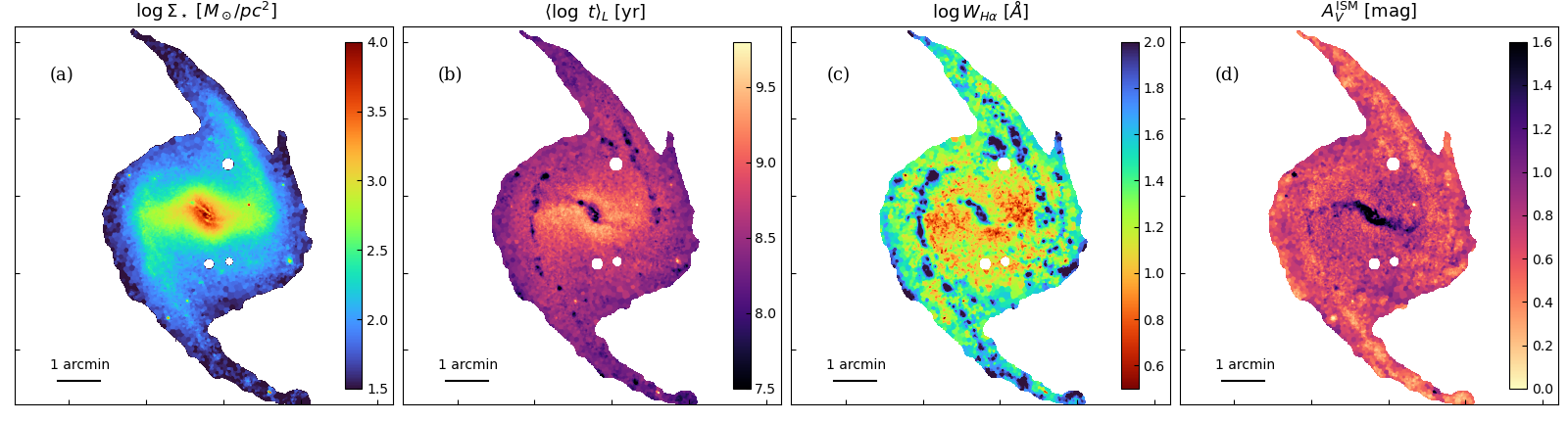}
      \caption{Example of IFS-like science with S-PLUS data for NGC 1365. (a) Stellar mass surface density ($\Sigma_\star$), (b) luminosity weighted mean log age ($\langle \log t \rangle_L$), (c) \Ha\ equivalent width ($W_{\Ha}$), and (d) stellar extinction ($A_V$) maps. 
      }
    \label{fig:NGC1365Maps_1}
\end{figure*}

\begin{figure*}
    \centering
    \includegraphics[width=\textwidth]{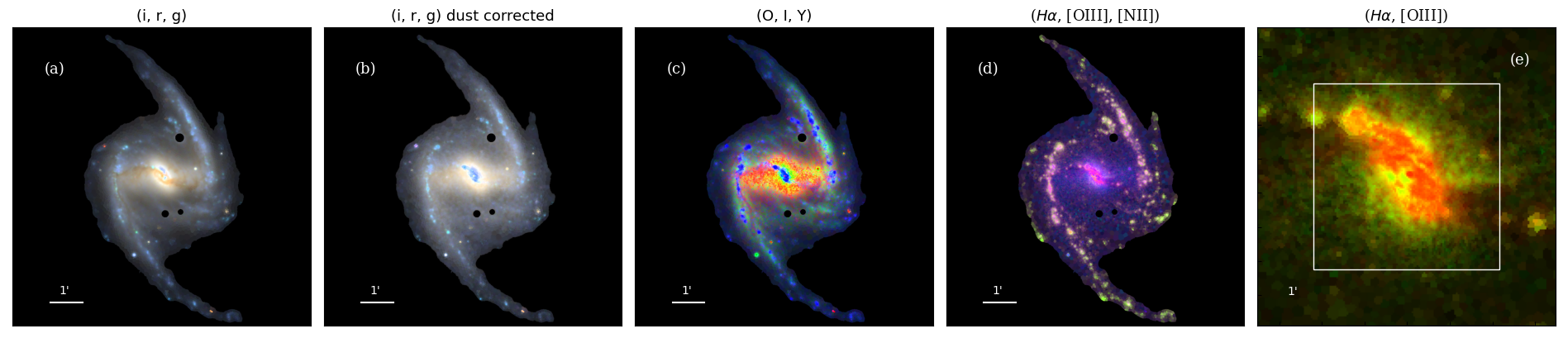}
    \caption{RGB composites combining respectively the (a) observed i, r, and g band fluxes, (b) the dust-corrected i, r, g fluxes, (c) the 5635 \AA\ continuum fluxes associated with old, intermediate, and young populations, (d) the \Ha, \oiii, and \nii\ fluxes. Panel (e) shows a map of \Ha\ (in R) and \oiii\ (G) of the central region, to be compared to the same map produced by \citealt{2018Venturi} (their figure 1) on the basis of MUSE data (field of view shown as a box).
    }
    \label{fig:NGC1365Maps_2}
\end{figure*}

\begin{figure*}
    \centering
    \includegraphics[width=\textwidth]{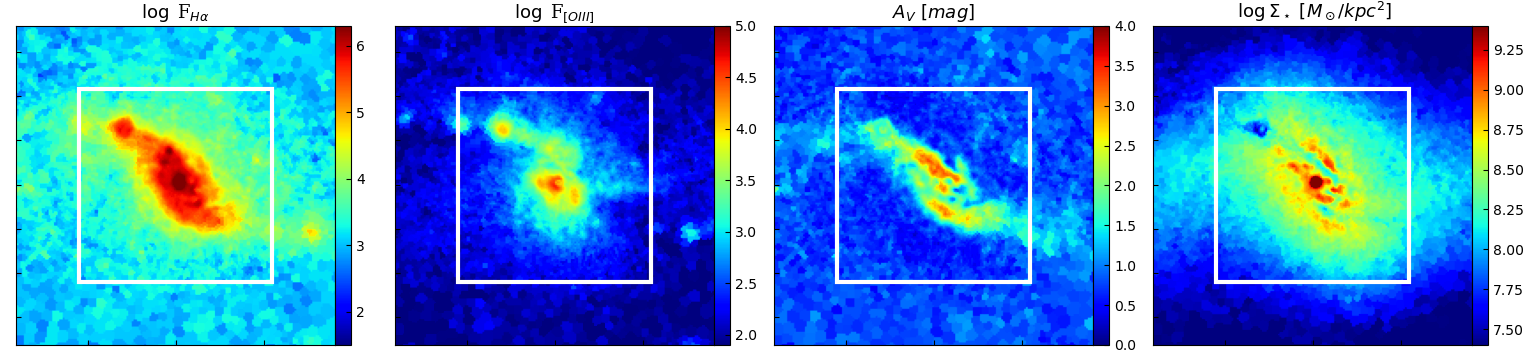}
    \caption{Zoom on the inner maps of $F_{\Ha}$, $F_{\oiii}$ (both in units of $10^{-20}$erg$\,$s$^{-1}\,$cm$^{-2}$, adjusted by the pixel scale difference between MUSE and S-PLUS), $A_V$ and $\Sigma_\star$ maps. The colour scales are matched to those of the same images in the \citealt{Gao2021MUSE}, who produced maps of these quantities derived on the basis of MUSE data over the inner $\sim 1 \times 1$ arcmin (marked by a box). 
    The $A_V$ map shown corresponds to the extinction applied to young stars and ELs (i.e. the sum of ISM and BC components), as this is more comparable to the \Ha/\Hb-based $A_V$ map of \citealt{Gao2021MUSE}. 
    }
    \label{fig:NGC1365Maps_3}
\end{figure*}

\begin{figure}
    \centering
    \includegraphics[width=\columnwidth]{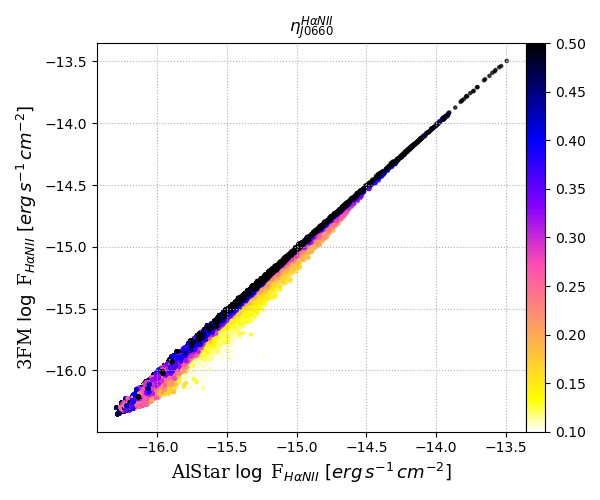}
    \caption{Comparison of the \nii6548  + \Ha\ + \nii6584 fluxes obtained by the 3 filters method (r, J0660, and i) and those from our analysis for 4431 Voronoi zones in NGC 1365. Points are colour-coded by fraction of the total J0660 flux that is due to these ELs.
    }
    \label{fig:3FM_x_AlStar}
\end{figure}

For nearby galaxies we can perform a spatially resolved analysis similar to that obtained with integral field spectroscopy (IFS). The two galaxies examined in integrated light in the previous section are examples of cases where a spatial analysis is clearly warranted. 

As a first experiment of IFS-like work with S-PLUS, let us examine the \alstar\ maps obtained for NGC 1365. The cube comprises $1100 \times 1100$ pixels ($10.08 \times 10.08$ arcmin). Besides a basic signal-to-noise spatial mask, we have further applied a low pass (Butterworth) filter and Voronoi binning (to reach a target $S/N$ of 20 in the five bluest filters combined). Details on these pre-processing steps will be presented elsewhere. The same stellar base described in the previous section was used. Also, the EL base and constraints are the same ones discussed in section \ref{sec:WHaN_based_constraints}.

Fig.\ \ref{fig:NGC1365Maps_1} illustrates the results. \alstar-based maps of the stellar mass surface density ($\Sigma_\star$), $\langle \log t \rangle_L$, $W_{\Ha}$, and $A_V^{\rm ISM}$ are shown in panels a--d, respectively. The spatial correspondence between regions of low age and large $W_{\Ha}$ (and vice-versa) is evident to the eye, illustrating that, without being forced to do so (which would be an acceptable strategy), the code identifies \hii\ regions both in the stellar and nebular components. The dust lane across the bar and nucleus is neatly visible in panel d. 
The age and dust maps indicate that star formation is plentiful in the central region, but not along the larger scale bar in which it is embedded.

Fig.\ \ref{fig:NGC1365Maps_2} shows some other products of our S-PLUS + \alstar\ analysis. Panel (a) shows an RGB composite based on the observed i, r, and g S-PLUS fluxes. Panel (b) show the same composite, but after correcting the fluxes by dust attenuation. Visibly, the bar-aligned dust lane disappears in the de-dusted image. Panel (c) shows the continuum fluxes (at our normalization wavelength of 5635 \AA) associated to old ($t > 1$ Gyr), intermediate (10 Myr--1 Gyr) and young ($\leq 10$ Myr) stellar populations, colour-coded onto R, G, and B channels, respectively. This higher order version of the $\langle \log t \rangle_L$ map in Fig.\ \ref{fig:NGC1365Maps_1} reveals a mixture of young and intermediate age populations in the central parts, the young stars in the spiral arms, and the predominantly old ones along the bar.

Fig.\ \ref{fig:NGC1365Maps_2}(d) shows an \Ha\ (in the R channel), \oiii\ (G), \nii\ (B) composite. \Ha\ dominates the emission in the central regions. This reflects both the star-forming activity in this region and the high dust content, which attenuates the  \oiii\ emission. \hii\ regions along the spiral arms also stand out clearly. The greener (stronger \oiii) colours of the outer \hii\ regions in this image are indicative of a negative nebular metallicity gradient, qualitatively consistent with what one expects for disc galaxies in general and for NGC 1365 in particular \citep{2022Chen}. The fainter emission (e.g., between the nucleus and the arms) have both lower $W_{\Ha}$ (panel c in Fig.\ \ref{fig:NGC1365Maps_1}) and larger \nii/\Ha\ (of order 0.5), consistent with the properties of diffuse ionized gas in spirals (e.g., \citealt{2018_Lacerda}, \citealt{2019Vale_Asari}).

The rightmost panel in Fig.\ \ref{fig:NGC1365Maps_2} shows \Ha\ (in the R channel) and \oiii\ (G) fluxes in the inner region of NGC 1365. This figure was made with the explicit goal of comparing it with the same map obtained from MUSE data by \cite{2018Venturi}. The image reveals both the circumnuclear star-formation, bright in \Ha, and the \oiii-bright biconical outflow originating from the Seyfert nucleus of this galaxy. Despite the huge differences in collecting aperture, exposure time, and spectral resolution, the S-PLUS+\alstar\ map in Fig.\ \ref{fig:NGC1365Maps_2}(e) looks like a coarser version of figure 1d of \cite{2018Venturi}, again illustrating the power of S-PLUS as an IFS-like machine for nearby galaxies.

This potential is further illustrated in Fig.\ \ref{fig:NGC1365Maps_3}, where we zoom in on the inner $\sim 1$ arcmin maps of the (from left to right) \Ha\ and \oiii\ fluxes, $A_V$, and $\Sigma_\star$ maps. These same properties were estimated on the basis of a full spectral analysis of MUSE data by \cite{Gao2021MUSE}. 
Comparing our results with theirs reveals a very good level of agreement, specially considering the differences in resolution.
The knot at the top left of the MUSE box in  Fig.\ \ref{fig:NGC1365Maps_3} corresponds to cluster ID 37 of the \cite{2023_Whitmore} study, which reports a log age/yr of 6.5 (6.6) and $A_V$ of 2.0 (2.6) mag with HST (JWST) data. These values compare well with our $\langle \log t \rangle_L=6.87$ and  $A_V=2.67$ mag.
These examples show that S-PLUS is capable of producing useful estimates of spatially resolved properties of both stellar population and ELs in the local Universe.

To close, let us compare our \alstar-based estimates of the total \Ha +\nii6548,6584 flux ($F_{\HaNii}$) with those based on the cruder, but more direct estimates obtained with the so called three filter method (3FM, \citealt{2015_Vilella_Rojo}).\footnote{Our implementation of this method includes the effects of $\nii\lambda\lambda6548, 6584$, \Ha, and $\sii\lambda\lambda 6717, 6731$ lines on the r, J0660, and i bands. It is thus somewhat more elaborate than that used in \cite{2015_Vilella_Rojo}, though, in practice, the two approaches yield very similar results.} Fig.\ \ref{fig:3FM_x_AlStar} compares the $F_{\HaNii}$ values obtained with the two methods for Voronoi zones in NGC 1365 (see Lopes et al.\ 2023, in prep., for more examples). The comparison is limited to the 4431 zones where the 3FM estimate of $F_{\HaNii}$ exceeds its uncertainty (i.e., $S/N > 1$).
Points are coloured by the fraction $\eta^{\HaNii}_{\rm J0660}$ of the total J0660 flux which comes from these three lines. 

The plot shows that the \alstar\ and 3FM estimates agree very well, with a median $\pm\, \sigma_{\rm NMAD}$ of the $\delta =$ \alstar\ minus 3FM values of $\delta \log F_{\HaNii} = 0.068 \pm 0.061$ dex. As seen in Fig.\ \ref{fig:3FM_x_AlStar}, the agreement is even better when ELs contribute more to the J0660 flux. For instance, we find $\delta \log F_{\HaNii} = 0.032\pm 0.021$ dex for zones where $\eta^{\HaNii}_{\rm J0660} > 30$\%. The difference for zones with small $\eta^{\HaNii}_{\rm J0660}$ is partly due to the effect of the \Ha\ absorption component (included in \alstar\ but not in the 3FM), which explains the somewhat larger values obtained by \alstar.

\section{Summary}
\label{sec:Conclusions}

S-PLUS is currently building a huge data set of galaxy photometry in its 7 narrow + 5 broad bands system. Transforming these 12 fluxes onto astrophysical information on stellar population and EL properties requires tools such as the \alstar\ code tested in this study.

After reviewing \alstar\ basics and introducing a novel semi-empirical strategy to improve its EL-estimation power (Section \ref{sec:AlStar}), we have applied the code to synthetic photometry of $\sim 10$k SDSS galaxies (Section \ref{sec:SDSS_simulations}), shifted to $z = 0.01$ to ensure the main optical ELs are sampled in S-PLUS narrow bands. Noise with a spectrum characteristic of S-PLUS was added to these data to achieve r-band signal-to-noise ratios typical of the envisaged near-Universe applications. The output stellar population and EL properties of these runs were then compared to those obtained by a previous detailed $\lambda$-by-$\lambda$ full spectral analysis (W19). The results of these comparisons may be summarized as follows:

\begin{enumerate}

    \item Stellar population properties recovered from the 12 S-PLUS bands compare well with those obtained from full spectral fitting. The statistics (median $\pm \sigma_{\rm NMAD}$) of $\delta =$ output $-$ input differences are $\delta \log M_\star = 0.04 \pm 0.06$ dex, $\delta \langle \log t \rangle_L = -0.03 \pm 0.16$ dex, $\delta \langle \log Z \rangle_L = 0.15 \pm 0.18$  dex and $\delta A_V = -0.03 \pm 0.14$ mag for $SN_r = 50$. 
    Even for the lowest $SN_r$ in our simulations we find acceptable differences given the huge compression in the observational input in  going from a full spectrum to its 12-bands S-PLUS representation. We conclude that S-PLUS is able to provide a useful first order characterization of stellar population properties even at relatively low $SN_r$. 

    \item 
    The main focus of this work was on ELs.
    We model their contribution to the S-PLUS photometry by complementing the stellar base with an EL base built to mimic ELs as found in real galaxies.  Stellar and EL components are fitted simultaneously. The specific EL base used in this work was built on the basis of the BPT diagram of SDSS galaxies. This scheme guarantees that the output EL properties are realistic (though not necessarily accurate).

    \item Confirming previous J-PLUS work, because they all fall under the J0660 filter, the sum of \Ha\ and \nii6548,6584 fluxes is always well recovered. Quantitatively, we retrieve their combined equivalent width ($\WHaNii)$ to within $\pm 0.02$ dex for our test sample, which (by design) spans from  weak ($W_{\Ha} = 1$ \AA) to  strong ($> 100$ \AA) EL systems.

    \item  The novelty comes in our ability to disentangle \Ha\ from \nii, as well as to estimate other ELs with remarkable precision for a photometric survey. This is achieved by imposing priors on the EL base guided by an initial (but robust) estimate of the value of $\WHaNii$. We use SDSS data to calibrate $\WHaNii$-based constraints on the \nii/\Ha\ and \oiii/\Hb\ ratios which, despite a few caveats, greatly improve our ability to recover EL properties beyond \Ha\ + \nii.

    \item  For the full 10k sample and $SN_r = 50$ we obtain $\delta = \log W$ is $0.04 \pm 0.16$, $0.04 \pm 0.08$, $0.13 \pm 0.23$, $0.01 \pm 0.05$, $0.01 \pm 0.08$, and $0.03 \pm 0.13$ dex for \oii, \Hb, \oiii, \Ha, \nii, and \sii, respectively. These statistics, which are already remarkably good, get even better for lines detected with $W > 5$ \AA.

\end{enumerate}

We have further presented example applications of the code to real S-PLUS data, including integrated light photo-spectra and a datacube (Section \ref{sec:Applications}). This second part of the paper showed the following:

\begin{enumerate}
    \item The method is capable of producing excellent fits to S-PLUS data. For the two whole-galaxy examples shown, the model matches the data to within 2.8\% (NGC 1365) and 1.7\% (NGC 1379). The derived stellar masses agree well with those obtained from independent photometry and a different method.  
    Because of their large angular extents, spectroscopic estimates of the stellar and EL properties of these two galaxies are available only for their central regions. Considering this caveat, as well as the differences in methodology and definitions, our rough comparison with literature results showed a reasonable level of agreement.

    \item One of the main applications envisaged for the method developed here is to obtain spatially resolved maps of stellar population and EL of nearby galaxies. A 
    $\sim 10 \times 10$ arcmin datacube for \mbox{NGC 1365} was analysed to showcase this kind of application. 
    The resulting maps trace very well the main stellar, dust and EL structures across this barred spiral. Comparison with previous MUSE-based properties derived for the inner regions reveals a very good level of agreement,  reinforcing the potential of S-PLUS+\alstar\ for IFS-like work. 
    
\end{enumerate}

Next steps include work both on the data and methodological fronts. 
Including more bands in the analysis (say, the GALEX NUV and FUV filters), for instance, is straight forward. 
Similarly, application to J-PAS photometry would both lead to better constraints and extend the redshift-range of applicability. 
Regarding the method itself, a promising way forward is to generalize the $\WHaNii$-based empirical prior used to improve the EL estimation. Given the plethora of correlations between observed properties of galaxies (e.g., \citealt{1997Worthey}, \citealt{2004Tremonti}, \citealt{1998Kennicutt}, \citealt{2022Quilley}), one can envisage a more-informative prior including, for instance, data on morphology or colours, to further improve the estimation of EL properties out of photometric surveys like S-PLUS.

\section*{Acknowledgements}

The S-PLUS project, including the T80-South robotic telescope and the S-PLUS scientific survey, was founded as a partnership between the Funda\c c\~ao de Amparo \`a Pesquisa do Estado de S\~ao Paulo (FAPESP), the Observat\'orio Nacional (ON), the Federal University of Sergipe (UFS), and the Federal University of Santa Catarina (UFSC), with important financial and practical contributions from other collaborating institutes in Brazil, Chile (Universidad de La Serena), and Spain (Centro de Estudios de F\'isica del Cosmos de Arag\'on, CEFCA). We further acknowledge financial support from the S\~ao Paulo Research Foundation (FAPESP), Funda\c c\~ao de Amparo \`a Pesquisa do Estado do RS (FAPERGS), the Brazilian National Research Council (CNPq), the Coordination for the Improvement of Higher Education Personnel (CAPES), the Carlos Chagas Filho Rio de Janeiro State Research Foundation (FAPERJ), and the Brazilian Innovation Agency (FINEP).

The authors who are members of the S-PLUS collaboration are grateful for the contributions from CTIO staff in helping in the construction, commissioning, and maintenance of the T80-South telescope and camera. We are also indebted to Rene Laporte and INPE, as well as Keith Taylor, for their important contributions to the project. From CEFCA, we particularly would like to thank Antonio Mar\'in-Franch for his invaluable contributions in the early phases of the project, David Crist\'obal-Hornillos and his team for their help with the installation of the data reduction package JYPE version 0.9.9, C\'esar \'I\~niguez for providing 2D measurements of the filter transmissions, and all other staff members for their support with various aspects of the project.
The authors also thank Ulisses Manzo Castello, Marco Antonio dos Santos, and Luis Ricardo Manrique for all their support in infrastructure matters.

JTB acknowledges a scholarship from FAPESC (CP 48/2021). FRH acknowledges funding for this work from FAPESP grant 2018/21661-9. LSJ acknowledges the support from CNPq (308994/2021-3) and FAPESP (2011/51680-6). AL acknowledges a postdoctoral fellowship from CONICET. AVSC acknowledges grants from CONICET and Agencia I+D+i.


\section*{Data Availability}

The complete S-PLUS DR4 catalogues and reduced fits images are available in the S-PLUS Cloud Database, at \url{https://splus.cloud}. The catalogues can also be accessed in Topcat using the TAP URL \url{https://splus.cloud/public-TAP/tap}, and through the Python package \texttt{splusdata} (see documentation and source code in \url{https://github.com/Schwarzam/splusdata}). The code with which the cubes were produced is available in \url{https://github.com/splus-collab/splus-cubes}.



\bibliographystyle{mnras}
\bibliography{MAIN.bib} 








\bsp	
\label{lastpage}
\end{document}
